\documentclass[12pt,letterpaper]{article}
\pdfoutput=1

\usepackage{graphicx,array}
\usepackage{mathrsfs} 
\usepackage{color}
\usepackage{latexsym}
\usepackage{amsthm}
\usepackage{amsmath}
\usepackage{amssymb}
\usepackage{dsfont}
\usepackage{epsfig}
\usepackage{slashed}
\usepackage{bbold}
\usepackage{psfrag}
\usepackage[svgnames]{xcolor}
\PassOptionsToPackage{caption=false}{subfig}
\usepackage{subcaption}
\usepackage{xfrac}
\usepackage{multirow}
\usepackage{booktabs}
\usepackage[colorlinks=true,linkcolor=MediumBlue,citecolor=Green,urlcolor=violet]{hyperref}
\usepackage{cite}
\usepackage[normalem]{ulem}
\usepackage{footnote}
\usepackage{pdflscape}
\usepackage[utf8]{inputenc}
\makesavenoteenv{tabular}
\makesavenoteenv{table}

\newcommand{\be}{\begin{equation}}
\newcommand{\ee}{\end{equation}}
\newcommand{\bea}{\begin{eqnarray}}
\newcommand{\eea}{\end{eqnarray}}

\def\mL{\mathcal{L}}
\def\mM{\mathcal{M}}

\def\mO{\mathcal{O}}

\newcommand{\LTW}[1]{{\color{red} \large\bf[#1]}}

\newcommand{\Eq}[1]{Eq.~(\ref{#1})}
\newcommand{\Ref}[1]{Ref.~\cite{#1}}

\newcommand{\olr}[1]{\overleftrightarrow{#1}}

\usepackage{terms}

\begin{document}

\title{Precision Measurement with Diboson at the LHC  }


\date{}

\author{Da Liu$^{a}$ and Lian-Tao Wang$^{b}$}

\maketitle

\begin{center}
\it {$^a$\,High Energy Physics Division, Argonne National Laboratory, Argonne, IL 60439}

\it {$^b$\, Department of Physics, Enrico Fermi Institute, and Kavli Institute for Cosmological Physics,University of Chicago, 5640 S Ellis Ave, Chicago, IL 60637, USA}
\end{center}

\begin{abstract}
Precision measurements at the LHC can provide probes of new physics, and they are complementary to direct searches. The high energy distribution of di-boson  processes ($WW,WZ,Vh$) is a promising place, with the possibility of significant improvement in sensitivity as the data accumulates. We focus on the semi-leptonic final states, and make projections of the reach for future runs of the LHC with integrated luminosities of 300 fb$^{-1}$ and 3 ab$^{-1}$. We emphasize the importance of tagging the polarization of the vector bosons, in particular for the $WW$ and $WZ$ channels. We employ a combination of kinematical distributions of both the $W$ and $Z$, and their decay products to select the longitudinally polarized $W$ and $Z$. We have also included our projections for the reach using the associated production of vector boson and the Higgs. We demonstrate that di-boson measurement in the semi-leptonic channel can surpass the sensitivity of the precision measurement at LEP, and they can be significantly more sensitive than the HL-LHC $h \to Z \gamma$ measurements. Compared with fully leptonic decaying $WZ$ channel, the reach from semi-leptonic channel can be better with effective suppression of the reducible background and systematic error. We  have also considered the reaches on the new physics mass scale in different new physics scenarios,  including the Strongly Interacting-Light Higgs (SILH), the Strongly Coupled Multi-pole Interaction (Remedios), and  the class of models with partially composite fermions. We find that in the SILH and non-compact Remedios scenario with large coupling $g_* > 7$, measurements in the di-boson channel is more sensitive than the Drell-Yan di-lepton channel at the HL-LHC. 
\end{abstract}

\section{Introduction}

Precision measurement at the LHC will be one of its most important legacies. Electroweak symmetry breaking is one of the central questions of the Standard Model. Focusing on electroweak sector of the Standard Model (SM), precision measurements can provide valuable lessons which will help us address this question. 

With the assumption that new physics particles would not be produced directly at the LHC, we parameterize their effect by a set of dimension 6 effective field theory (EFT) operators \cite{Grzadkowski:2010es,Elias-Miro:2013mua,Contino:2013kra}. In this paper, we focus on operators relevant to
electroweak precision measurements. Such measurements have been carried out at LEP \cite{ALEPH:2010aa}, with typical precisions on the order of $10^{-3}$.  This can be interpreted as constraining the scale of new physics to be higher than $ \Lambda \sim$2 TeV. At the LHC, effects of new physics can potentially grow with energy. For example, if the leading effect is through interference  between dim-6 operator and the SM, it could grow with energy as  $\propto E^2/\Lambda^2 $.  In this case, since energies around TeV can be probed at the LHC, we only need a $20 \% $ measurement to achieve a reach similar to that of LEP precision measurements. In order to fully take advantage of this effect, it is important to focus on final states whose amplitude not only grows like $E^2$, but also interferes with a approximately constant SM amplitude. In practice, this requires carefully designed cuts to select such final states. As we review in Section~\ref{sec:general}, in the two-vector-boson channels ($WW$ and $WZ$), an obvious channel would be the production of  longitudinally polarized vector bosons. Polarization tagging would be crucial to separate it from channels with other polarizations. At the same time, such inference is guaranteed for the $Vh$ channel. 

The  present bounds on these operators from LHC di-boson processes have been studied in \Ref{Ellis:2014dva,Falkowski:2014tna,Falkowski:2015jaa,Butter:2016cvz,Zhang:2016zsp,Falkowski:2016cxu,Green:2016trm,Biekoetter:2014jwa,Baglio:2017bfe,Ellis:2018gqa}. The prospects of probing these operators in the tri-lepton channel,  $WZ \to 3 \ell \nu$ and the di-lepton channel $WW\to 2\ell 2\nu$ has been studied in \Ref{Franceschini:2017xkh,Chiesa:2018lcs}, while the Higgs associated production channels  for the SM case have also been considered in \Ref{Butterworth:2015bya,Tian:2017oza}. In this paper, we focus our attention on the semi-leptonic channel of $WW, \ WZ $ production. In comparison with the pure leptonic channel, the semi-leptonic channel has larger rate. At the same time, it presents new challenges.  Not being able to clearly distinguishing hadronically decaying $W$ and $Z$, we will have to consider them together.  Unlike the $WZ$ channel, $WW$  channel does not have the sharp ``amplitude zero" feature in the central region. In this paper, we employ additional information from the distribution of the decay products of the vector boson to help tagging its polarization. We have also included our analysis for the $Vh$ channel, which are in broad agreement with the results in Ref.~\cite{Franceschini:2017xkh}. Based on these analysis, we make projections for the sensitivity to new physics. 

The rest of this paper is organized as follows. In Section~\ref{sec:general}, we describe the EFT framework of our analysis, and offer general discussions of key aspects in the analysis of di-boson channels. We present our analysis of the potential of the semi-leptonic channel, which is the main result of this paper, in Section~\ref{sec:semileptonic}. In Section~\ref{sec:reachnp}, we apply the result of our analysis to estimate reaches in new physics scale in  several more specific scenarios. Our conclusions are contained in Section~\ref{sec:conclusion}. 

\section{General Considerations}
\label{sec:general}
After integrating out new physics, the SM Lagrangian is modified by the addition of higher dimensional operators. We have 
\beq
\mathscr{L}=\mathscr{L}_{\rm SM} + \sum_{i} \frac{c_i}{\Lambda^2} \mO_i + \cdots 
\eeq
where $\Lambda$ has no $\hbar$ dimension and it should be interpreted as a mass threshold. We have only included dimension 6 operators. 
The operators most relevant for the di-boson channel are
\beq
\begin{split}
&{\cal O}_W  =\frac{ig}{2 }\left( H^\dagger  \sigma^a \overleftrightarrow{D}^\mu H \right )D^\nu  W_{\mu \nu}^a , \qquad {\cal O}_B =\frac{ig'}{2 }\left( H^\dagger  \overleftrightarrow{D}^\mu H \right )\partial^\nu  B_{\mu \nu} ,\\
&\mO_{2W} = - \frac12 D^\mu W_{\mu\nu}^a D_\rho W^{a\rho\nu}, \qquad \mO_{2B}= - \frac12 \partial^\mu B_{\mu\nu} \partial_\rho B^{\rho\nu},\\
&{\cal O}_{HW} = i g(D^\mu H)^\dagger\sigma^a(D^\nu H)W^a_{\mu\nu}, \qquad
{\cal O}_{HB}= i g'(D^\mu H)^\dagger(D^\nu H)B_{\mu\nu},\\
&\mathcal{O}_{3W} = \frac{1}{3!} g \epsilon_{abc} W_\mu^{a\nu}W^b_{\nu\rho}W^{c\rho\mu}, \qquad {\cal O}_T = \frac{g^2}{2}(H^\dagger \olr{D}^\mu H )(H^\dagger \olr{D}_\mu) H, \\
&\mO_R^u = i g^{\prime 2} \left( H^\dagger  \overleftrightarrow{D}_\mu H \right ) \bar{u}_R \gamma^\mu u_R, \qquad \mO_R^d = i g^{\prime 2} \left( H^\dagger  \overleftrightarrow{D}_\mu H \right ) \bar{d}_R \gamma^\mu d_R,\\
&\mO_L^q = i g^{\prime 2} \left( H^\dagger  \overleftrightarrow{D}_\mu H \right ) \bar{q}_L \gamma^\mu q_L, \qquad \mO_L^{(3)q} = i g^2 \left( H^\dagger\sigma^a  \overleftrightarrow{D}_\mu H \right ) \bar{q}_L\sigma^a \gamma^\mu q_L,
\end{split}
\label{eq:d6ops}
\eeq
where $H^\dagger \olr{D}_\mu H  \equiv H^\dagger D_\mu H - (D_\mu H)^\dagger H$. 
From this list, we will not further consider  $T$-parameter operator ${\cal O}_T$.  It has been well constrained by LEP experiment, and it is unlikely that LHC measurement can reach a comparable level.  We have also not included the operator ${\cal O}_H = \frac{1}{2f^2} (\partial |H|^2 )^2$ in the list. It modifies the  Higgs gauge boson coupling. Current results of Higgs coupling  measurement have already constrained  $f \gtrsim 800$ GeV, and the precision can reach $f \gtrsim 1200$ GeV with HL-LHC. It will lead to strong $W_LW_L(hh)$ scattering, as dictated by the Goldstone Equivalent Theorem~\cite{Chanowitz:1985hj,Wulzer:2013mza}. However, the effect is more prominent at higher energies $\sim (f/v)^2\times$TeV. The sensitivities of LHC to ${\cal O}_H$ in the di-boson channels  are  weak, reaching $f \gtrsim 350$ GeV at the HL LHC in the double Higgs final states~\cite{Contino:2010mh}  and $f \gtrsim 550$ GeV at the HL LHC in the same sign di-lepton channel of $W^\pm W^\pm$~\cite{Szleper:2014xxa}. It can't compete with Higgs coupling measurement. 

The contributions of these operators to scattering amplitudes depend on the final states.  We will consider the so called di-boson processes $q \bar{q} \to V_1 V_2$ and  $q \bar{q} \to Vh$, where $V = W^{\pm}, Z$.  With our normalization, the largest SM amplitude is a constant\footnote{There are also t-channel poles in the forward region, here for simplicity we are focusing on the central region.}. For dim-6 operators, their contributions to the amplitudes can grow at most as $E^2$. Hence, we should look for a channel with interference between the SM and new physics amplitude grows as $E^2$, or at least grows with energy. In order to have the energy growing behavior, it is not enough to just have the contribution of dim-6 operators to the amplitude to grow with energy. It is crucial to have the corresponding SM amplitude not decreasing at least as fast with energy. This condition can in principle be relaxed if the SM background interfering with the signal is the only SM background. In this case, we can have good sensitivity as long as $S/\sqrt{B}$ grows with energy. A SM background decrease with energy can in principle satisfy this condition, even if the interference piece does not grow. However, in practice, such cases are difficult to find. There are almost always (ir)reducible SM backgrounds which do not decrease with energy. Hence, the channels which have interference piece growing with energy remain our best hope. 

\begin{table}[h!]
\begin{center}
\caption*{ $q_L \bar{q}_R \rightarrow W^+ W^-$}
\begin{tabular}{|c|c|c|c|c|c|c|c|}
\hline
$(h_{W^+}, h_{W^-})$  &  SM  & $\mO_W$ & $\mO_{HW}$ & $\mO_{B}$ &  $\mO_{HB}$ & $\mO_{3W}$ \\ 
  \hline
    \hline
$(\pm,\mp)$         & $1$     &  0    &  0    &  0  &0  &  0   \\ 
\hline
$(0,0)$         & $1$      &  $\frac{E^2}{\Lambda^2}$ &  $\frac{E^2}{\Lambda^2}$ &  $\frac{E^2}{\Lambda^2}$ & $\frac{E^2}{\Lambda^2}$  &0  \\ 
\hline
$(0,\pm),(\pm,0)$         & $\frac{m_W}{E}$      &  $\frac{E m_W}{\Lambda^2} $  &  $\frac{E m_W}{\Lambda^2} $ &  $\frac{E m_W}{\Lambda^2} $ &$\frac{E m_W}{\Lambda^2}$ &  $\frac{E m_W}{\Lambda^2}$  \\ 
  \hline
$(\pm,\pm)$         & $\frac{m_W^2}{E^2}$      &  $\frac{m_W^2}{\Lambda^2}$  &  $\frac{m_W^2}{\Lambda^2}$ &   $\frac{m_W^2}{\Lambda^2}$  &0&  $\frac{E^2}{\Lambda^2}$  \\ 
\hline
\end{tabular}
\bigskip
\caption*{ $q_R \bar{q}_L \rightarrow W^+ W^-$}
\begin{tabular}{|c|c|c|c|c|c|c|}
\hline
$(h_{W^+}, h_{W^-})$  &  SM  & $\mO_W$ & $\mO_{HW}$ & $\mO_{B}$ & $\mO_{HB}$  & $\mO_{3W}$ \\
  \hline
    \hline
$(\pm,\mp)$         & $0$     &  0    &  0    &  0    & 0 &  0   \\ 
\hline
$(0,0)$         & $1$      &  $\frac{m_W^2}{\Lambda^2}$ &    $\frac{m_W^2}{\Lambda^2}$  &  $\frac{E^2}{\Lambda^2}$ & $\frac{E^2}{\Lambda^2}$ &  0   \\ 
\hline
$(0,\pm),(\pm,0)$         & $\frac{m_W}{E}$      &  $\frac{m_W^2 m_Z^2}{\Lambda^2 E^2} $  & $\frac{E m_W}{\Lambda^2}$ &  $\frac{E m_W}{\Lambda^2}$ & $\frac{E m_W}{\Lambda^2}$  &  $\frac{m_W^2 m_Z^2}{\Lambda^2 E^2}$ \\
  \hline
$(\pm,\pm)$         & $\frac{m_W^2}{E^2}$      &  $\frac{m_W^2}{\Lambda^2}$  &  $\frac{m_W^2}{\Lambda^2}$  &  $\frac{m_W^2}{\Lambda^2}$ & 0&$\frac{m_W^2}{\Lambda^2}$  \\ 
\hline
\end{tabular}\end{center}
\caption{\label{tab:WW} High energy behaviour for the helicity amplitudes $q \bar{q} \rightarrow W^+ W^-$, where we omit the gauge couplings $g^2, g^{\prime 2}$  in front of the amplitudes~\cite{Hagiwara:1986vm}. $\mO_{2W,2B}$ has similar behaviour as $\mO_{W,B}$.  $E$ can be thought as half of the partonic center of mass energy (i.e. the energy of single $W$ boson). The zeros in the table mean that  there are no such amplitude contribution at all in the zero mass limit of the quarks. For the $WZ$, the only non-zero amplitudes are for the left-handed quarks and only $\mO_{W, HW}$ operators have energy growing behaviour in the purely longitudinal helicity state. }
\end{table}

Perhaps the most straightforward cases to consider are the $Wh$ and $Zh$ channels. In this case, new physics amplitude interfere with the full Standard Model amplitude. The only challenge would be to identify the final states amid the reducible SM backgrounds. This has been demonstrated to be feasible~\cite{Butterworth:2008iy}. In particular, boost technologies play an important role in separating signal from reducible background. At the same time,  the boosted regime is also precisely the place for enhancing the new physics effect. Further studies of this channel have been presented recently~\cite{Butterworth:2015bya,Tian:2017oza}.

The channels with two vector gauge bosons are more complicated.  
From Table~\ref{tab:WW} (see also~\cite{Azatov:2016sqh}), we conclude that the most promising channels are those with longitudinally polarized vector bosons, as the interference piece grows with energy as $\propto E^2$. 
Hence, we expect isolating events with longitudinal polarized vector bosons will be particularly important. There can be two strategies in achieving this goal. One is to take advantage of the fact that final state with different polarizations have different kinematical distribution~\cite{Hagiwara:1986vm}. A particularly useful example is the so-called ``amplitude  zero" in the transversely polarized $WZ$ final states~\cite{Baur:1994ia,Frye:2015rba}. In this case, using kinematical cuts which select the central region enhances the longitudinally polarized component. This approach has been used in Ref.~\cite{Franceschini:2017xkh}. 

The second strategy is directly tagging the polarization of a gauge boson from  the angular distribution of its decay products. Such a polarization tagging can be challenging. The basic difference would be in the angular distribution of the decay product in the rest frame of the gauge boson. Even with perfect reconstruction and identification, one would not expect the difference between different polarizations to be much more than order one. In practice, one strategy would be to reconstruct the rest frame of the gauge boson, and use the angular distribution of the decay product~\cite{Han:2009em}. The systematically error in the reconstruction needs to be taken into account. Another strategy would be to use the kinematical feature of the decay product in the lab frame. This has the advantage of skipping the step of reconstructing the rest frame of the gauge boson. However, some of the information of the angular distribution will be washed out. 

\begin{savenotes}
\begin{table}[ht]
\begin{center}
\begin{tabular}{|c|c|}
\hline
  Observable& $\delta O /  O_{\text{SM}}$  \\ 
  \hline
    \hline
$W_L^+ W_L^-$  & $  \left[(c_W + c_{HW} - c_{2W})T_f^3  + (c_B + c_{HB} - c_{2B})   Y_f t_w^2 \right]   \frac{E^2}{\Lambda^2}, \ c_f \frac{E^2}{\Lambda^2} $  \\
\hline
  $W_T^+ W_T^-$   & $c_{3W} \frac{m_W^2}{\Lambda^2} + c_{3W}^2\frac{E^4}{\Lambda^4}, \ c_{TWW } \frac{E^4}{\Lambda^4}$ \\
    \hline
$W_L^\pm Z_L$  & $  \left(c_W + c_{HW} - c_{2W}+ 4 c_L^{(3)q} \right)  \frac{E^2}{\Lambda^2} $  \\ 
\hline
$W_T^\pm Z_T(\gamma)$   & $c_{3W} \frac{m_W^2}{\Lambda^2} + c_{3W}^2\frac{E^4}{\Lambda^4},\ c_{TWB} \frac{E^4}{\Lambda^4}$ \\
    \hline
$W_L^\pm h$  & $  \left(c_W + c_{HW}  - c_{2W}+ 4 c_L^{(3)q}  \right)  \frac{E^2}{\Lambda^2} $  \\
\hline
 $Z h$ & $  \left[(c_W + c_{HW} - c_{2W})T_f^3  - (c_B + c_{HB} - c_{2B})   Y_f t_w^2 \right]   \frac{E^2}{\Lambda^2},\ c_f  \frac{E^2}{\Lambda^2}$  \\
  \hline
$Z_T Z_T$  & $( c_{TWW } + t_w^4  c_{TBB } - 2 T_f^3 t_w^2 c_{TWB})\frac{E^4}{\Lambda^4} $  \\ 
\hline
 $\gamma\gamma$   &$( c_{TWW } +   c_{TBB } + 2 T_f^3  c_{TWB})\frac{E^4}{\Lambda^4}$  \\
 \hline
$\hat{S}$&       $(c_W + c_B) \frac{m_W^2}{\Lambda^2}$  \\
\hline
$h \rightarrow Z \gamma$   & $(c_{HW} - c_{HB})\frac{(4\pi v)^2}{\Lambda^2}$  \\
\hline
  $h\rightarrow W^+ W^-$&  $(c_W + c_{HW}) \frac{m_W^2}{\Lambda^2}$  \\
\hline
\end{tabular}
\vspace{0.5cm}
\caption{ Observables for probing the higher dimensional operators. $c_f$ denotes the Wilson coefficients of the fermionic operators in \Eq{eq:d6ops}. For reference, we have also included contributions from potential dim-8 operators with Wilson coefficients denoted by $c_{TX}$. See Appendix C of \Ref{Liu:2016idz} for the definition of the dimension-8 operators. }
\label{tab:obs}
\vspace{0.5cm}
\end{center}
\end{table}
\end{savenotes}

A list of diboson channels and other observables,  and the contributions from new physics operators,  are presented in Table~\ref{tab:obs}. For reference, we have also included the contribution of dim-8 operators, where we refer to Appendix C of \Ref{Liu:2016idz} for the definitions.  We see that each of the observables receive contributions from multiple operators. More specifically, the contributions to di-boson production in the high energy limit depend on the following combinations~\cite{Franceschini:2017xkh}:
\beq
\begin{split}
c_{q_L}^{(3)} &= c_W + c_{HW}  - c_{2W}+ 4 c_L^{(3)q},  \\
c_{u_L}^{(1)} &= c_B + c_{HB} - c_{2B}  + 4 c_L^q,  \\
c_{d_L}^{(1)} &= c_B + c_{HB} - c_{2B}  - 4 c_L^q,  \\
c_{u_R}^{(1)} &= c_B + c_{HB} - c_{2B} +3 c_{u_R}, \\
c_{d_R}^{(1)} &= c_B + c_{HB} - c_{2B} - 6c_{d_R}. \\
\end{split}
\label{eq:comb}
\eeq
This result can be understood easily by using the following operator relations (together with additional equations of motion) to rewrite  operators $\mO_{HW,HB}, \mO_{W,B}, \mO_{2W,2B}$ in terms of the operators with more fields, such as  $\mO_{WW,WB,BB},\mO_{L,R}^f,\mO_{4f},\mO_{y_f}$, and so on~\cite{Elias-Miro:2013mua,Contino:2013kra}.
\bea
\label{eq:op_relation}
{\cal O}_B&=&{\cal O}_{HB}+ {\cal O}_{BB}+\frac14{\cal O}_{WB}\,, \quad
{\cal O}_{WB}= gg' (H^\dagger \sigma^a H) W^a_{\mu\nu}B^{\mu\nu}\,,\quad
{\cal O}_{BB}=g^{\prime 2} H^\dagger H B_{\mu\nu} B^{\mu\nu}, \nonumber\\
{\cal O}_W&=&{\cal O}_{HW}+ \frac14{\cal O}_{WW}+\frac14{\cal O}_{WB}\, ,\qquad  {\cal O}_{WW}=g^2 H^\dagger H W^a_{\mu\nu} W^{a\mu\nu}\,.
\eea
The resulting set of operators are called the Warsaw basis [1]. For example, from the first relation on the first line of  Eq.~(\ref{eq:op_relation}), the operators $\mO_B$ and $\mO_{HB}$ contribute in the same way to longitudinal di-boson final states, since $\mO_{WB}$ and $\mO_{BB}$ only contribute to the production of the transverse di-boson final states. Similarly, the operator $\mO_B$ can be related to the $\mO_{L,R}^f$ operators by equations of motion of the hyper-charge gauge field.

It is impossible to distinguish separate contributions from operators within each combination from di-boson measurement. Besides di-boson production, one of the most important observable is the oblique $S$-parameter~\cite{Peskin:1991sw}, which has been well constrained by LEP precision electroweak measurement~\cite{ALEPH:2010aa}.  It depends on a different combination of the operators $\mO_W + \mO_B$. Therefore, it is complementary to the di-boson measurement at the LHC. At the same time, we do not expect large cancellation among operators short of large fine-tuning or special symmetry. In this case, we can view the LEP measurement of the $S$-parameter as setting a generic limit on size of $\mO_W$ and $\mO_B$, and use that as a target for the LHC experiments. Similar argument also applies to the measurement of Higgs rare decay $h \to Z \gamma$ at the HL-LHC, which will be sensitive to the operator combination $\mO_{HW} - \mO_{HB}$.  For this measurement at the HL-LHC, we will use the projections made in Ref.~\cite{Dawson:2013bba}.

So far, our discussion is at the level of parton level cross section. The observable cross section is obtained after convolution with parton distribution functions. Taking this into account, the signal cross section scales with energy as
\be
\sigma_{\rm sig} \propto {\cal{M}}_{\rm SM} \left( \frac{E}{\Lambda} \right)^{d-4} \left(\frac{1}{E} \right)^{n_L+2},
\ee
where $d$ is the dimension of the EFT operator responsible for the signal, and ${\cal{M}}_{\rm SM} $ is the SM amplitude with which the new physics amplitude interferes. $n_L$ parameterizes the dependence of parton luminosity on the parton center of mass energy. Parton luminosity is a sharp falling function of $E$. Typically, $n_L$ is a large power, around $4$ - $6$. If the search channel is statistics dominated, we have
\be
\frac{S}{\sqrt{B}} \propto  \left( \frac{E}{\Lambda} \right)^{d-4} \left(\frac{1}{E} \right)^{n_L/2+1} \times \sqrt{{\cal{L}}},
\ee
where ${\cal{L}}$ is the integrated luminosity. To obtain this qualitatively scaling behavior, we have made the crude approximation that $\sigma_{\rm bkg} \sim |{\cal{M}}_{\rm SM}|^2$. This means the sensitivity of different energy bin depends on the dimension of the EFT operator to be probed. For example, for $d=6$ operators in Eq.~(\ref{eq:d6ops}), lower energy bins have higher sensitivity. On the other hand, for probing $d=8$ EFT operators, we expect higher energy bins yield better sensitivity. However, the assumption of statistics domination is certainly not realistic. Systematical error is very important particularly for precision measurements. Lower energy bins, typically with a smaller $S/B$, will be more affected (and sometimes dominated) by systematics. Therefore, in reality, the most sensitive energy bin is typically determined by a trade off between systematics and statistics.


\section{Semi-leptonic channel of di-bososn processes}
\label{sec:semileptonic}
We will focus on the following semi-leptonically decaying channels at the LHC:
\beq
\begin{split}
&p p  \rightarrow  W V \rightarrow \ell \nu q \bar{q} , \qquad \text{BR}(W^+W^-\rightarrow \ell \nu q \bar{q}) = 29.2\%, \qquad \text{BR}(W^\pm Z\rightarrow \ell \nu q \bar{q}) = 15.1\%\\
&p p  \rightarrow  W h \rightarrow \ell \nu b \bar{b} , \qquad \text{BR} = 12.6\%\\
&p p  \rightarrow  Z h  \rightarrow \ell^+ \ell^- b \bar{b} , \qquad \text{BR} = 3.92\%\\
&p p  \rightarrow  Z h  \rightarrow \nu \bar{\nu}  b \bar{b}, \qquad \text{BR} = 11.6\% \\
\end{split}
\eeq
where $\ell = e, \mu$ and $V=W,Z$. We have listed the branching ratios of the semileptonic final states under consideration. 

For the Monte Carlo simulation, we first implement the dimension-six operators in \Eq{eq:d6ops} in an UFO model by using FeynRules~\cite{Alloul:2013bka}. We then use  MadGraph5~\cite{Alwall:2014hca} to simulate the signal and background events at LO. The cross sections of the processes considered in this paper is also calculated using MadGraph5 at the LO. For the studies in this paper, we have used NNPDF 2.3LO1~\cite{Ball:2012cx} as the parton distribution functions.

\subsection{$WV$ processes}
\begin{figure}[h!]
\begin{center}
\includegraphics[width=0.45\textwidth]{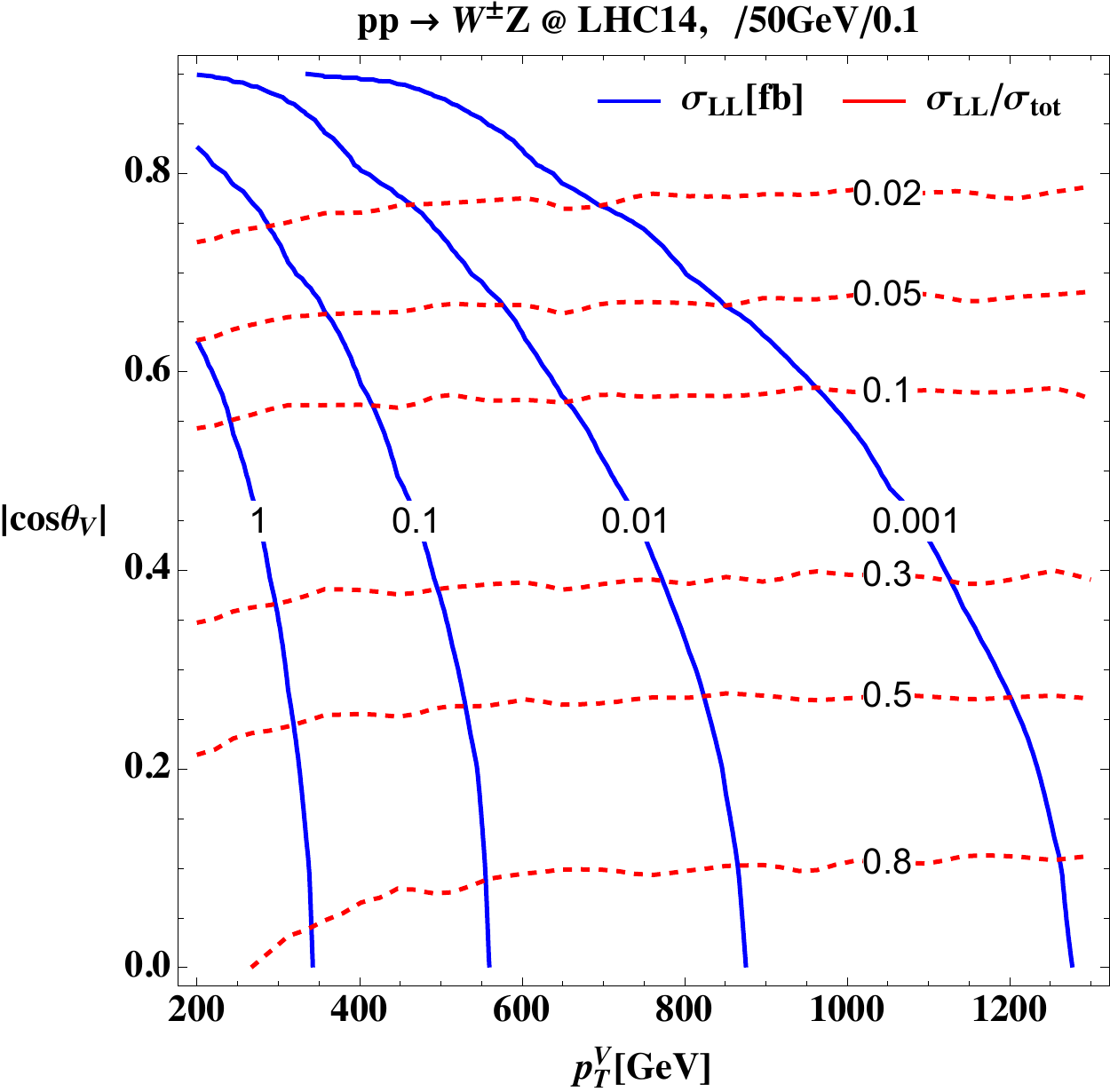}
\includegraphics[width=0.445\textwidth]{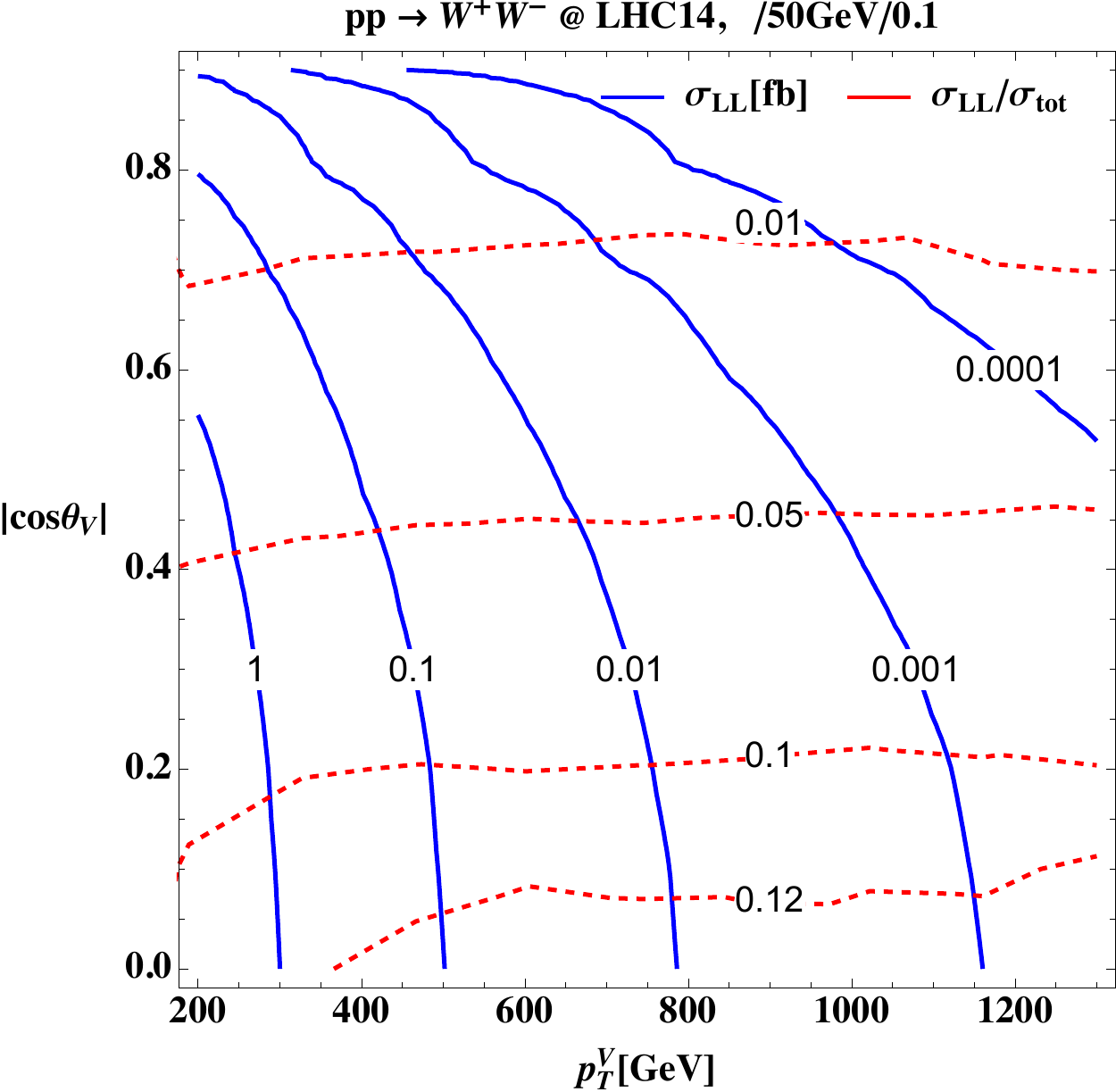}\\
\end{center}
\caption{Contours of production cross section of longitudinally polarized vector bosons $\sigma_{LL}$ and its ratio to total cross section, $\sigma_{LL}/\sigma_{tot}$,  in the $|\cos\theta_V| - p_{T}^V$ plane. $\theta_V$ is the scattering angle in the parton center-of-mass frame.  Left (right) panel is for $WZ$ ($WW$) production. We require $|\eta_{V}| < 2.5$. }
\label{fig:ptcostheta}
\end{figure}

We start from the semi-leptonic final states from the $WV$ processes. The longitudinal modes of $WV$ tend to be produced more centrally than the transverse ones. Two possible kinematical variables which can capture this feature are the transverse momentum, $p_T^V$, and the scattering angle in the parton center of mass frame, $\theta_V$, of the vector bosons. In Fig.~\ref{fig:ptcostheta}, we plotted the contours of the production cross section of longitudinally polarized vector bosons $\sigma_{LL}$, and its ratio to the total cross section,  $\sigma_{LL}/\sigma_{tot}$,  in the $|\cos\theta_V| - p_{T}^V$ plane.
We see that the $W_LZ_L$ can be dominant in the central region, while $W_LW_L$ is at most $10\%$ of the total rate. This is due to the presence (absence) of the so called ``amplitude zero" in $WZ$ ($WW$) channels~\cite{Baur:1994ia}. 
The behavior of the contours  can be understood qualitatively.  In the high energy regime, we can approximately neglect effects of the gauge boson masses $m_{W,Z}$. The differential production cross section for vector bosons with helicity $h_{V1}$ and $h_{V2}$ from initial parton $i$ and $j$ is:

\beq
\frac{d^2\sigma^{h_{V1}h_{V2}}}{dp_T^V d\cos\theta_V}  = \frac{1}{128\pi N_c} \sum_{ij}  \frac{\beta}{E^2}\frac{dL_{ij} }{dE} \frac{dE}{dp_T^V}\times |\mM^{h_{V1}h_{V2}}_{ij}(\theta_V, E)|^2
\eeq
where $E$ is the energy for single vector boson in the partonic center-of-mass frame. We have:
\beq 
p_T^{V} = p\sin\theta_V, \qquad \beta = \frac{p}{E}, \qquad \frac{d E}{dp_T^V} = \frac{p_T^V}{E \sin^2\theta_V} = \frac{\beta}{\sin\theta_V}.
\eeq
with $p = |\vec{p}|$ denotes the magnitude of the three-momentum of the gauge boson in the partonic-center-of-mass frame. 
For simplicity, we define the helicity states in the partonic-center-of mass frame. $\frac{dL_{ij} }{dE}$ is the parton luminosity defined as:
\beq
\frac{dL_{ij} }{dE} = \frac{8E}{S}\int_{s/S}^1 \frac{dx}{x}f_i(x,\mu) f_j(s/Sx,\mu), \qquad s = 4 E^2.
\eeq
where $s$ denote the square of partonic center-of-mass energy and $S$ means the proton-proton center-of-mass energy square. 
First, we consider  the $WW$ production. To get a qualitative understanding, we can ignore the contribution from hypercharge gauge coupling since it is small in comparison with the $SU(2)_L$ contribution. In the high-energy limit, we have
\beq
|\mM^{TT}_{ WW}|^2 = \frac{g^4}{32}\left(\frac{s^2}{t^2} + \frac{s^2}{u^2}\right)\sin^2\theta_V (1 + \cos^2\theta_V), \qquad |\mM^{LL}_{WW}|^2 = \frac{g^4}{32} \sin^2\theta_V ,
\eeq
where the amplitudes are summed over initial states $u\bar{u}$ and $\bar{u}u$. These are even functions of $\cos\theta_V$. Including the contribution from $d\bar{d} + \bar{d}d$ does not change the form of the squared amplitudes. Thus, the parton luminosity  can be factored out, and the ratio  ${d^2\sigma^{LL}_{WW}/}{d^2 \sigma^{TT}_{WW}}$ only depends on the ratio of the squared amplitudes
\beq
\frac{d^2\sigma^{LL}_{WW}}{d^2 \sigma^{TT}_{WW}} \sim   \frac{1} {8} \frac{(1 - \cos^2\theta_V)^2}{(1 + \cos^2\theta_V)^2}. 
\label{eq:Wratio}
\eeq
Since the total cross section is dominated by the production of the transversely polarized $W$s, Eq.~(\ref{eq:Wratio}) explains the flat contours for this ratio in the right panel of Fig.~\ref{fig:ptcostheta} in the large $p_T$ regime.  The factor $1/8$ in front of the right hand side of Eq.~(\ref{eq:Wratio}) also explains the small value $\sim 0.1$ in the most central region with $\cos\theta_V \rightarrow 0$. 

A very similar analysis applies to $WZ$ except that there is an amplitude-zero for the transverse mode production in the central region.  More specifically (again neglecting the contribution from the hypercharge), the squared amplitudes for the production of longitudinally and transversely polarized modes are
\beq
|\mM^{TT}_{ WZ}|^2 = \frac{g^4}{32}\left(\frac{s}{t} - \frac{s}{u}\right)^2\sin^2\theta_V (1 + \cos^2\theta_V), \qquad |\mM^{LL}_{WZ}|^2 = \frac{g^4}{16} \sin^2\theta_V.
\label{eq:WZratio}
\eeq
As in the case of $WW$ production, the squared amplitudes  are the same for initial states $u\bar{d} + \bar{d} u$ and $d\bar{u} + \bar{u}d$. Eq.~(\ref{eq:WZratio}) then explains the flat behavior for the contours of the ratio in the left panel of Fig.~\ref{fig:ptcostheta}.  In contrast to the $WW$ channel,  here the transverse amplitude vanishes in the $\cos\theta_V \sim 0$.  The ratio of the polarized production cross section is
\beq
\frac{d^2\sigma^{LL}_{WZ}}{d^2 \sigma^{TT}_{WZ}} \sim   \frac{1} {8\cos^2\theta_V} \frac{1 - \cos^2\theta_V}{1 + \cos^2\theta_V} 
\eeq
where it is clear that the longitudinal component is dominant in the central region, also shown in Fig.~\ref{fig:ptcostheta}.
 Since it can be challenging to fully distinguish hadronic $W$ and $Z$ at the LHC, both signal and background will receive contribution from both $WW$ and $WZ$ channels. Therefore, event selection based on simple kinematical cuts such as  $p_{T}^V$ and $\cos \theta_V$ will always suffer from the contamination from the transversely polarized $W$s and may not achieve optimal results. 
 
\begin{figure}[h!]
\begin{center}
\includegraphics[width=0.48\textwidth]{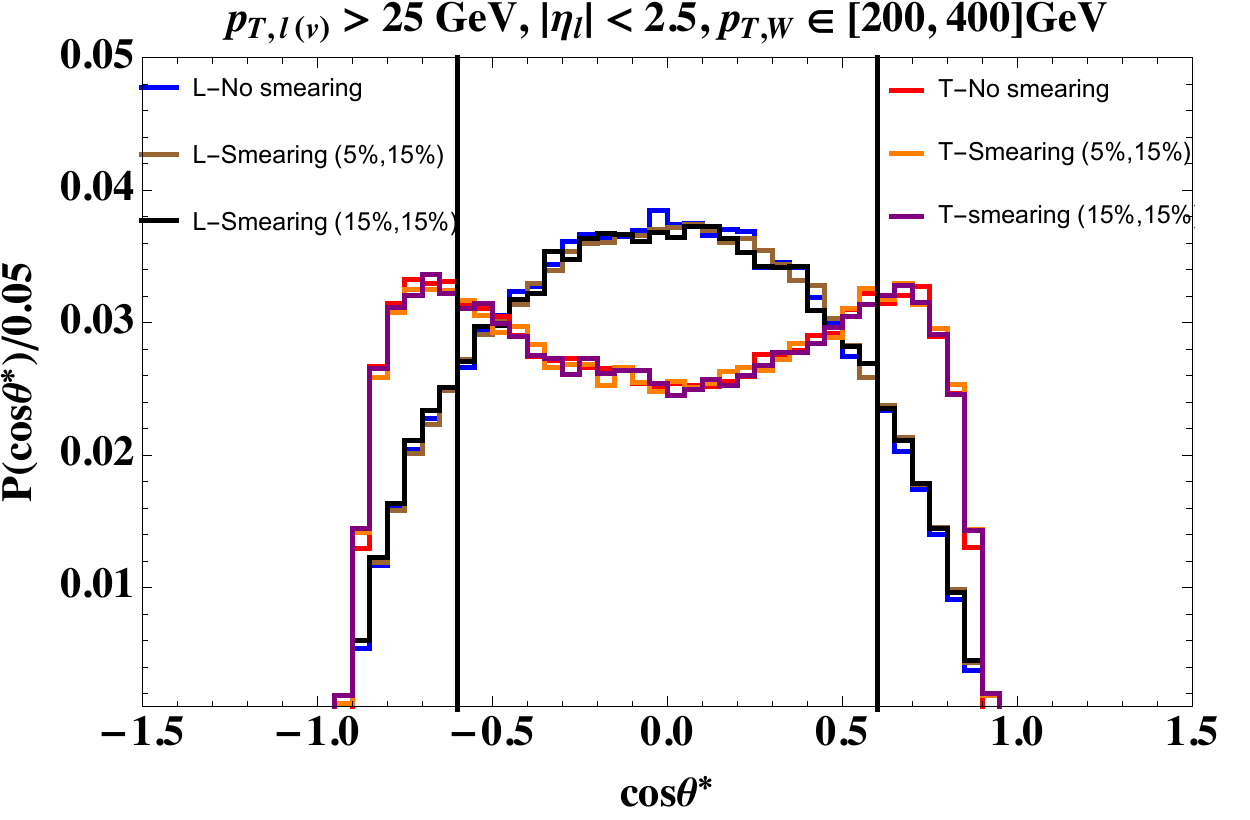}
\includegraphics[width=0.48\textwidth]{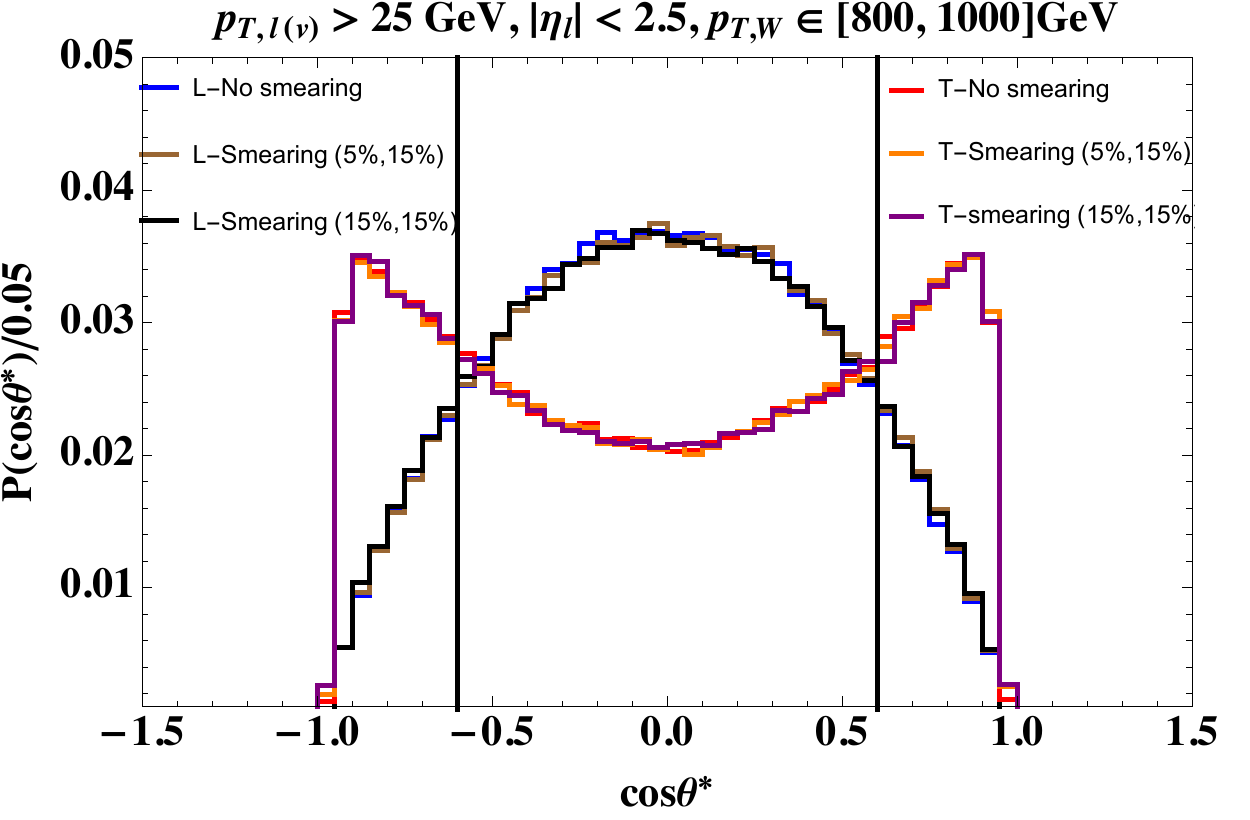}
\end{center}
\caption{Distribution of the $\cos\theta^*$ for the longitudinal  $W$  and transverse $W$-bosons  with $p_{T,W} \in [200,400]$ GeV (left  plot) and  $p_{T,W} \in [800,1000]$ GeV (right plot). The transverse $W$ bosons include both $+$ and $-$ helicities. The distributions are normalized to one. The blue and the red lines are  for the truth-level leptons and neutrinos, obtained from MadGraph5 \cite{Alwall:2014hca} LO simulation. The brown and orange lines are obtained by smearing the truth-level energy of leptons by 5\% and neutrino by 15\%.  The black and purple lines are obtained by smearing the truth-level energy of  both leptons and neutrino by 15\%. }
\label{fig:polartag}
\end{figure}

For the semi-leptonic  channel, polarization tagging using the information of the decay products can provide additional information to further enhance the signal. Such a strategy has been considered in Ref.~\cite{Han:2009em}. Here, we 
further explore its use in the case under consideration. The basic strategy is based on the well-known results that the  distribution of the polar angle $\theta^*$  for the lepton in the $W$-rest frame is different for longitudinally and the transversely polarized $W$ bosons.  The $z-$axis is typically chosen as the direction of the momentum of the $W$-boson in the laboratory frame~\cite{Han:2009em}. The probability  distributions of the polar angle for different helicity states in the $W^+$ decay are given by (see Appendix~\ref{app:Wpolar})
\beq
P_+ = \frac38 (1 - \cos\theta^*)^2, \qquad P_- = \frac38 (1 + \cos\theta^*)^2, \qquad P_0 = \frac34 \sin^2\theta^*
\label{eq:polar}
\eeq
Note that $\cos\theta^*$ can be obtained directly from the momenta of the lepton and neutrino in the laboratory frame  as\footnote{ In practice, the transverse momentum of the neutrino is identified with the missing energy. The longitudinal momentum of the neutrino is obtained by imposing the mass shell conditions for the neutrino and the $W$ boson.}  (see Appendix~\ref{app:Wpolar} for a more detailed derivation):
\beq
\cos\theta^* = \frac{E_\ell - E_\nu}{|\vec{p}_\ell +\vec{p}_\nu|}.
\label{eq:thetastar}
\eeq
Normalized distributions of reconstructed $\cos \theta^* $ from longitudinally and transversely polarized $W$s are shown in Fig.~\ref{fig:polartag} \footnote{Note that for the quarks from hadronically decaying $W$ boson, the distribution is the same. However, it is not possible to construct similar simple observables since we can not identify the charge of the quark very well. Instead, some jet substructure variables need to be used to take advantage of this information. }.  A major uncertainty in reconstructing the rest frame of the W boson is the detector resolution in measuring the momenta of its decay products. As an example, we can use the CMS detector performance during LHC Run 1~\cite{GoyLopez:2016trw}.  For the electrons with  $p_T \sim $45 GeV,  the energy resolution is better than 2\% in the central region  ($|\eta| <$ 0.8), and is 2\%-5\% elsewhere. For the muons, the energy resolution is 1.3 - 2.0\% in the barrel and better than 6\% in the endcaps in the $p_T$ region of [20,100] GeV. For the high $p_T$ muons, the resolution in the barrel is better than 10\% up to 1 TeV. The jet energy resolution is approximately given by the following formula:
\beq
\frac{\Delta E}{E} \approxeq \frac{100\%}{\sqrt{E[\GeV]}} \oplus 5\%
\eeq 
Usually, the transverse missing energy resolution is dominated by the hadronic activity of the event. Similar results can be found for the ATLAS detector~\cite{Aad:2009wy}. To estimate the resolution effects on the $\cos\theta^*$ distribution,  we included the  Gaussian smearing of lepton and neutrino energy scale with  following two benchmark  resolutions:
\beq
\begin{split}
&(1) \qquad \frac{\Delta E_\ell}{E_{\ell}} = 5\%, \qquad \frac{\Delta E_\nu}{E_{\nu}} = 15\%,\\
&(2) \qquad \frac{\Delta E_\ell}{E_{\ell}} = 15\%, \qquad \frac{\Delta E_\nu}{E_{\nu}} = 15\%
\end{split}
\eeq
where the second benchmark can be thought as the smearing effects in the hadronically decaying $W$ boson.
 From the plots, we can see that the distribution is relatively stable under such smearing. This is due to the fact that $\cos \theta^*$ is reconstructed as a ratio, as shown in Eq.~(\ref{eq:thetastar}). 


\begin{table}[!t]
\begin{center}
\caption*{$p_{T,W} \in [200,400]$ GeV}
\begin{tabular}{|c|c|c|c|c|c|c|c|c|c|}
\hline
Cut& $|\eta_{W,Z}| < 2.5$& $p_{T,\ell(\nu)} > 25\text{ GeV}, |\eta_\ell| < 2.5$ &  $|\cos\theta^*| < 0.6$ & $\epsilon_{p_T,\eta} \times \epsilon_{\cos\theta^*}$  \\
  \hline
 efficiency L      &0.572    &0.943   & 0.810 & 0.764 \\
  \hline
 efficiency T      &0.572    &0.776   &  0.665 & 0.516\\
\hline
\end{tabular}
\end{center}
\begin{center}
\caption*{$p_{T,W} \in [800,1000]$ GeV}
\begin{tabular}{|c|c|c|c|c|c|c|c|c|}
\hline
Cut& $|\eta_{W,Z}| < 2.5$& $p_{T,\ell(\nu)} > 25\text{ GeV}, |\eta_\ell| < 2.5$ &  $|\cos\theta^*| < 0.6$  &$\epsilon_{p_T,\eta} \times \epsilon_{\cos\theta^*}$\\
  \hline
 efficiency L      &0.853   &0.995 & 0.791 &0.787   \\
  \hline
 efficiency T      &0.854    &0.921  &0.553 & 0.509  \\
\hline
\end{tabular}
\end{center}
\caption{ Two benchmarks for the longitudinal and transverse poarization tagging. The transverse $W$ bosons include both $+$ and $-$ helicities with equal probability.}
\label{tab:polareff}
\end{table}

From Fig.~\ref{fig:polartag} and \Eq{eq:thetastar}, we see that the decay products are more central (forward) for longitudinally (transversely) polarized $W$s. 
This implies that the energies of the lepton and the neutrino in the lab frame tend to be symmetric for the longitudinally polarized $W$ boson. On the other hand,  the decay products of $W$s with transverse polarization are more asymmetric. One of them tends to be hard, while the other tends to be soft.  Due to these kinematical differences, the $p_T, \eta$ cut on the charged leptons will already have some differential power on the longitudinal and transverse $W$s. In addition, we can impose a cut on the reconstructed $\cos \theta^*$ directly to further distinguish the two polarizations. In Table~\ref{tab:polareff}, we have presented the effect of these cuts in two different kinematical regimes,  one with moderately boosted $W$-boson $p_{T,W} \in [200, 400]$ GeV,  and the other with  highly boosted $W$ boson $p_{T,W} \in [800, 1000]$ GeV. Table~\ref{tab:polareff} shows that $\cos\theta^*$ cuts can help with the signal significantly  for highly boosted region. For the moderately boosted region, $p_T, \eta$ cuts on leptons are already quite useful in suppressing the contamination  from transverse $W$s. The addition of a cut on $\cos\theta^*$ does not significantly improve it.  Based on this discussion, in the following analysis, we will use the following  values for the polarization tagging:
\beq
\label{eq:polartag}
\epsilon_L  \equiv \epsilon^L_{p_T, \eta} \times  \epsilon^L_{\cos\theta^*} = 0.75, \qquad \epsilon_T \equiv \epsilon^T_{p_T, \eta} \times  \epsilon^T_{\cos\theta^*} = 0.5.
\eeq

The difference in the distribution of decay products also has a direct impact on tagging the hadronically decaying $W$-boson using the jet substructure method, with the longitudinal $W$-tagging efficiency higher by~40\% (see Ref.~\cite{Khachatryan:2014vla}).  One can also use jet substructure observables to develop a polarization tagger based on these kinematical features. We will leave this interesting topic for a future study. For this moment, we will assume the same polarization tagging efficiencies for the hadronically decaying $W,Z$ gauge bosons as \Eq{eq:polartag}.\footnote{For the reducible backgrounds, we have assumed same polarization tagging efficiencies as \Eq{eq:polartag}. This is fine for the leptonically decaying $W$ boson in the $W$+jets background because of its transverse nature. For the QCD jet faking the hadronically decaying $W,Z$ bosons, it remains to be seen to what extent some jet shape variable carrying information of $\cos\theta^*$ could help.  The assumption does not affect  our second benchmark in \Eq{eq:benchmark}, as we assume that the reducible backgrounds are negligible.  }

\begin{figure}[h!]
\begin{center}
\includegraphics[width=0.5\textwidth]{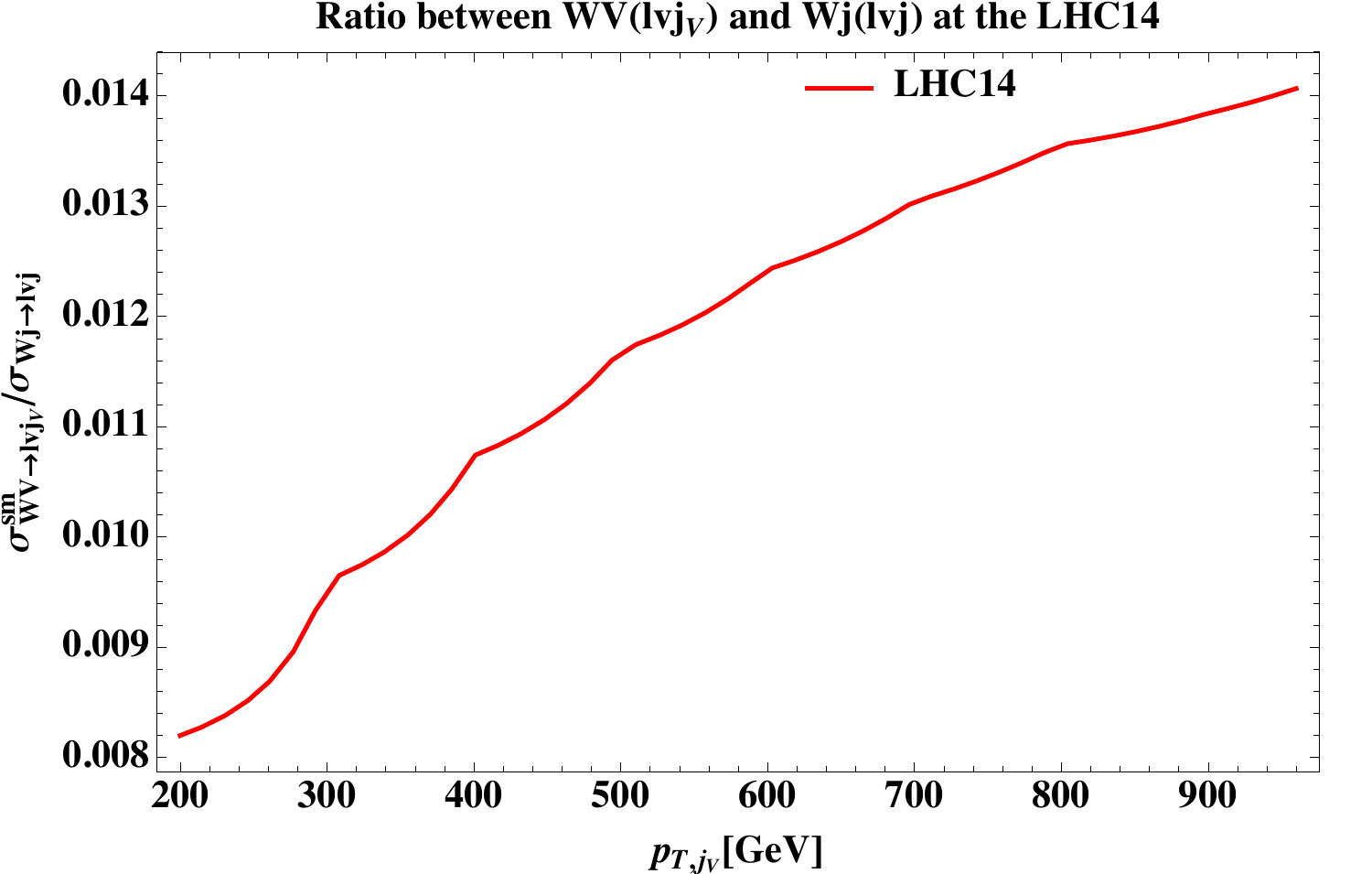}
\end{center}
\caption{The cross section ratio between SM $WV\rightarrow \ell \nu j_V$  and  $Wj\rightarrow \ell \nu j$  at the 14 TeV LHC. The branching ratios to the particular final states are taken into account. We have required $|\eta_{W,j}| < 2.5$. }
\label{fig:SoverB}
\end{figure} 

 The dominant reducible background for the semi-leptonic channel is expected to be $W+$jets, as shown in the 8 TeV analysis~\cite{Sirunyan:2017bey,Aaboud:2017cgf}.
 We show in Fig.~\ref{fig:SoverB} the  LO cross section ratio between the  SM $WV\rightarrow \ell \nu j_V$  and  $Wj\rightarrow \ell \nu j$ as a function of $p_{T}^V$ at the 14 TeV LHC, where $j_V$ denotes the jet resulting from hadronic decay of a vector boson. The simulation is carried out  at the parton level using MadGraph~\cite{Alwall:2014hca}, and we have required $|\eta_{V,j}| < 2.5$. We see that this ratio ranges from $1\%$ to $1.5\%$ as $p_T^{\rm jet}$ increases from 200 GeV to 1 TeV.
The most important tool to suppress this background is tagging  hadronically decaying $W,Z$ using jet substructure observables. 
 In Ref.~\cite{ATLAS:2017jiz}, the ATLAS collaboration has studied the performance of the $W$-boson tagging in Run 2, and made projections of the efficiency of $W$-tagging and the rejection of the QCD-jet background.  A benchmark point in the $p_T^{\rm jet}$ range [500, 1000] GeV for the $W$-tagging efficiency is $\epsilon^{\rm tag}_{W} = 0.3$, while the miss-tagging efficiency for QCD-jet is $\epsilon_{j}^{\rm miss} = 0.004$. Combining this with the cross section ratio shown in Fig.~\ref{fig:SoverB},  we could suppress the reducible background $W$+jets to the same order of SM $WV$ production in the semi-leptonic channel.  Ref.~\cite{ATLAS:2017jiz} doesn't show the results for the $W$ tagging efficiency below 0.3. In  an earlier study of \Ref{ATL-PHYS-PUB-2015-033}, the ATLAS collaboration has shown the $W$-tagging efficiency below 0.3, but with higher overall mistagging efficiency for the QCD-jet. For the $\epsilon_W^{\rm tag} = 0.3$, the miss-tagging rate is  $\epsilon_{j}^{\rm miss} = 0.006$, while for  $\epsilon_W^{\rm tag} = 0.1$, the miss-tagging rate  is $\epsilon_{j}^{\rm miss} = 0.0014$. Compared with \Ref{ATL-PHYS-PUB-2015-033}, \Ref{ATLAS:2017jiz} has improved  the QCD jet mis-tagging rate by $33\%$ for the $\epsilon_W^{\rm tag} = 0.3$.  If we assume the same improvement can be  achieved for the case of $\epsilon_W^{\rm tag} = 0.1$, the mis-tagging rate for QCD jet becomes $\epsilon_{j}^{\rm miss} = 0.0009$. The resulting reducible background for $WV$ channel is roughly $20\%$ of the SM $WV$ process and thus is sub-dominant. In our study, we will assume that for $\epsilon^{\rm tag}_W = 0.1$, the reducible backgrounds can be reduced to a negligible level.
 We choose the following two benchmarks for the performance of vector boson tagging. 
\beq
\begin{split}
\label{eq:benchmark}
&\epsilon^{\rm tag}_V = 0.3, \qquad n_{\rm red} = n_{\rm irred} \\
&\epsilon^{\rm tag}_V = 0.1, \qquad n_{\rm red} = 0 \\
\end{split}
\eeq
where $V$ denotes the hadronically decaying $W,Z$ bosons. $n_{\rm irred}$ is the number of irreducible background events, which comes from SM $WV$ production. $n_{\rm red}$ is the number of reducible background events which mostly comes from SM $W$+jets production.  We have assumed that the tagging efficiencies for hadronically decaying $Z$s and $W$s are similar.

To summarize, the cross section  in the semi-leptonically decaying channel from $WV$ production is given by
\beq
\begin{split}
\sigma_{\rm semi-lep} &=\sum_{p,p^\prime = L,T} \sigma^{pp^\prime}_{WW} (p_{T,W} > 200\GeV, |\eta_W| < 2.5)\times BR_{WW\rightarrow\ell\nu jj} \times \epsilon_V^{tag}  \times \epsilon^{p}\times \epsilon^{p^\prime} \\
&+ \sum_{p,p^\prime = L,T}  \sigma_{WZ}^{pp^\prime}  (p_{T,V} > 200\GeV, |\eta_V| < 2.5) \times BR_{WZ\rightarrow\ell\nu jj} \times\epsilon_V^{tag}  \times \epsilon^{p}\times \epsilon^{p^\prime} ,
\end{split}
\eeq
with various efficiencies taking on benchmark values discussed in this section. 

\subsection{$Vh$ production}
For the $Vh(b\bar{b})$ processes, the longitudinal component is dominant in the high-energy region for the SM. Therefore, we would not need to worry about contamination from final states with transverse polarization. 
 In this case, suppressing the reducible background is essential to enhance the new physics effects. The dominant reducible backgrounds  are $Vb\bar{b}, t\bar{t}$, and single top processes.  It has been firmly established that the use of jet substructure method  can be  effective in separating signal from background in the kinematical regime where Higgs has a sizable boost~\cite{Butterworth:2008iy,Butterworth:2015bya}.  This is also the regime where new physics effects considered here are enhanced. In particular, Ref.~\cite{Butterworth:2015bya} studied the prospect for the discovery of  the SM-like Higgs using boosted Higgs tagging method,   mainly in the $Wh \rightarrow \ell \nu b \bar{b}$ channel. They demonstrated that, in the kinematic region $p_{T}^V > 200$ GeV, a signal to background ratio of $S_{\rm SM}/B_{\rm red} \sim 0.2$ is achievable.  Here, $S_{\rm SM}$ refers to the rate of SM $Wh$ associated production, while $B_{\rm red}$ is the rate of the reducible background. The signal efficiency  obtained by the analysis using jet substructure in Ref.~\cite{Butterworth:2015bya}  depending on the $p_T$ bins. For the bins [200,400], [400,600], and  $> 600$ GeV, the efficiencies  are  0.1, 0.2, and 0.3, respectively.\footnote{The number of events  in each bin is given by: $
n^i = \sigma \times \text{BR}\times \epsilon^i_{\rm tot} $.} More recently, Ref.~\cite{Tian:2017oza} has studied this SM processes in the 0, 1, and 2-lepton states at the 13 TeV LHC using a combination of  boosted Higgs tagging variables.   They obtained $S_{\rm SM}/B_{\rm red} \sim 1$ with signal efficiency $\epsilon_{\rm tot} \sim 0.1$  in the kinematic region $p_{T}^V > 200$ GeV. Of course, such phenomenological studies of the performance of the Higgs taggers and background rejection power are not fully realistic, they will need to be further studied by the experimental collaborations. At the same time, we also expect potential improvement both on the optimization of the variables and the reduction experimental systematics. In our projection for the potential of HL-LHC, we will use the following benchmark:
\beq
\label{eq:vhbenchmark}
\epsilon_{\rm tot} = 0.1, \qquad n_{\rm red} = n_{\rm irred}
\eeq
in the 0, 1, and 2 -lepton channels of $Vh$ production,  focusing on the boosted regions $p_{T,V} > 200$ GeV. Here, $n_{\rm irred}$ refers to the number of events from the SM $Vh$ production.

\begin{figure}[!tb]
\begin{center}
\includegraphics[width=0.49\textwidth]{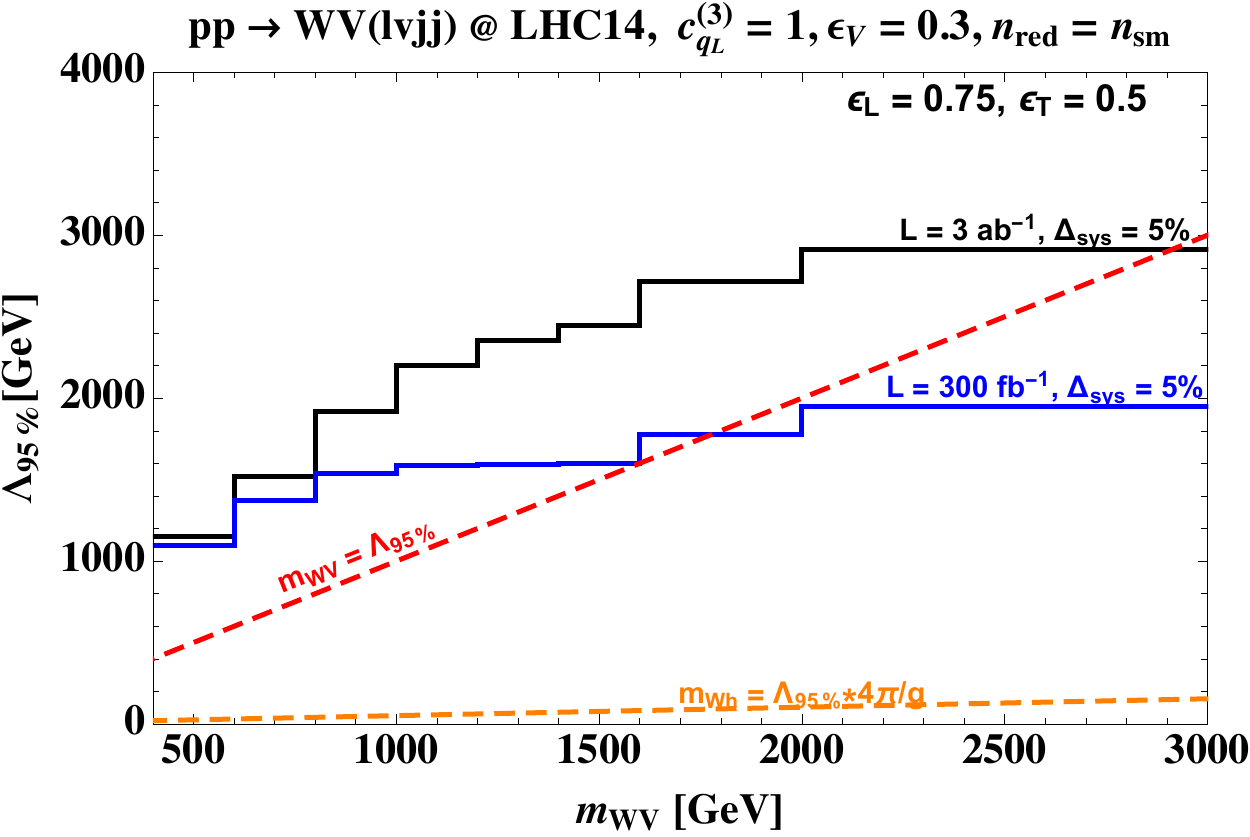}
\includegraphics[width=0.49\textwidth]{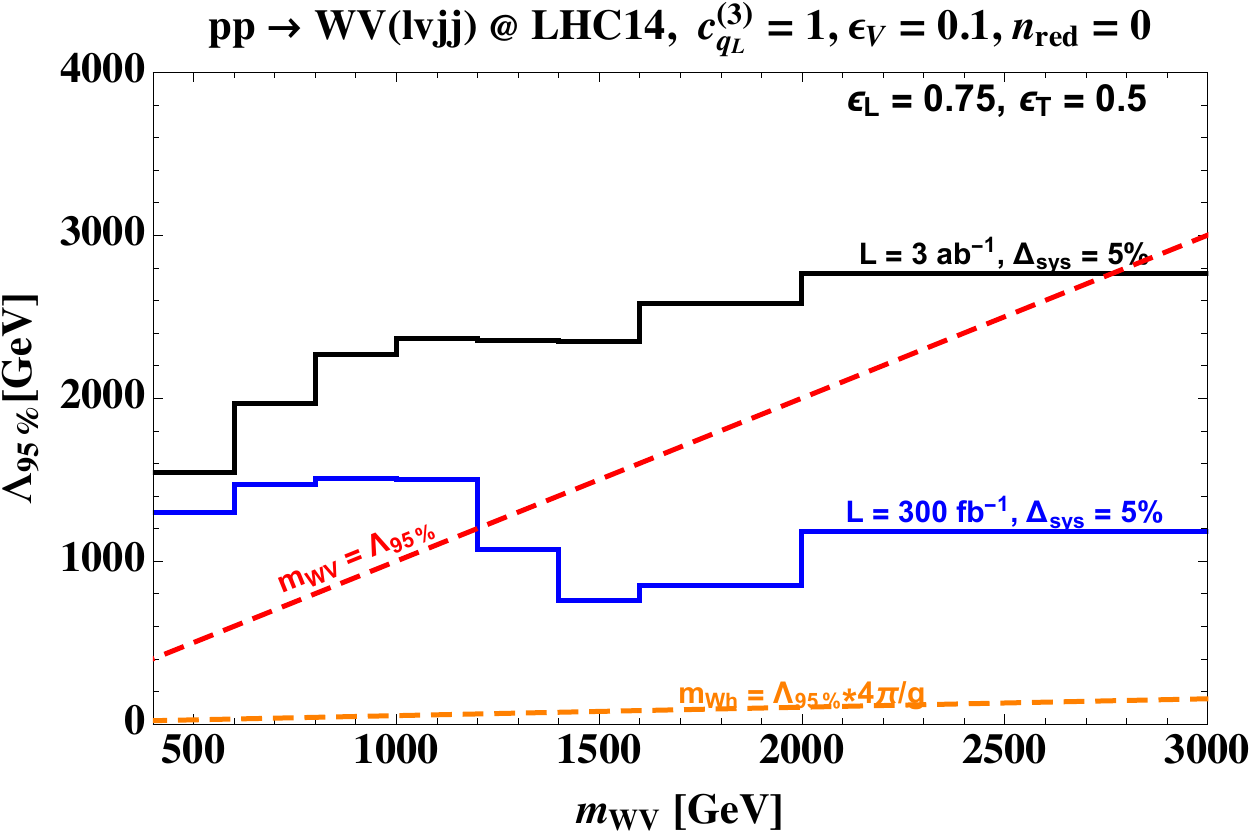}\\
\includegraphics[width=0.49\textwidth]{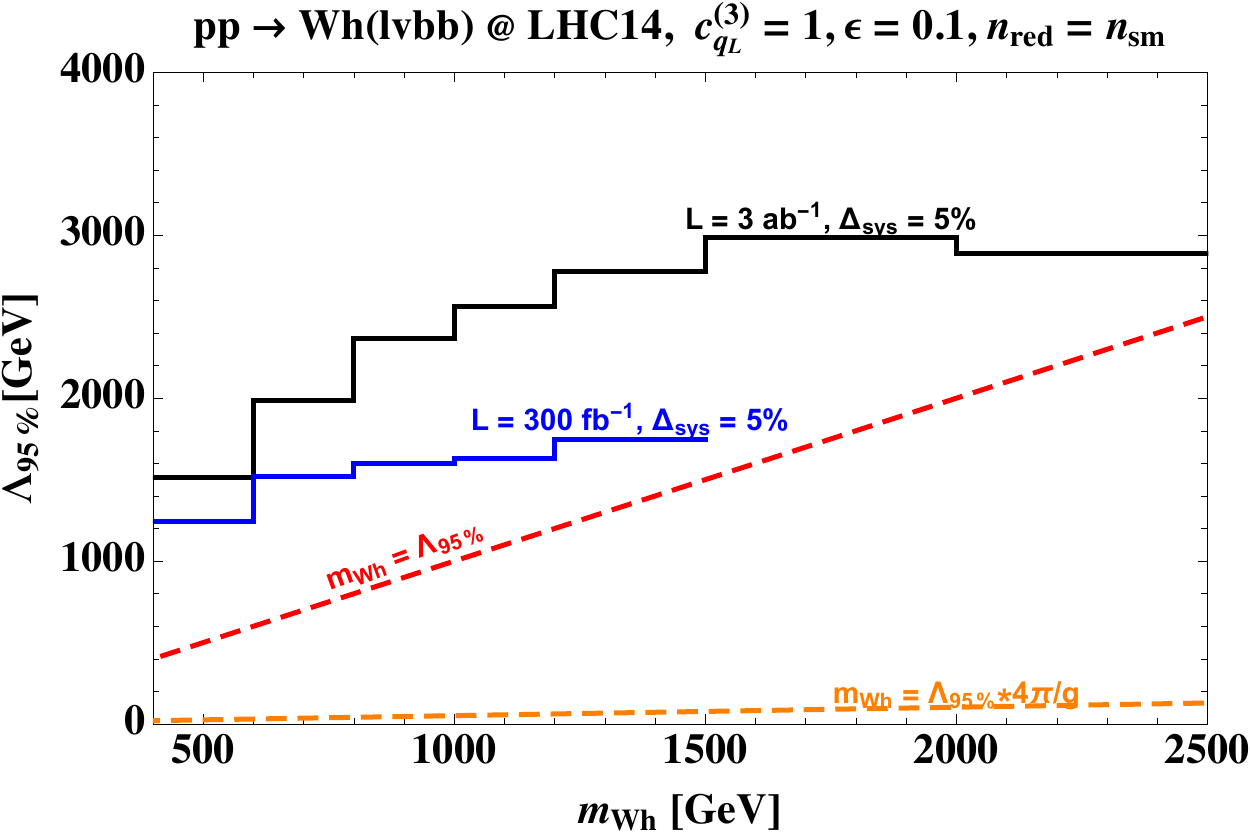}
\end{center}
\caption{$\Lambda_{95\%}$, the 95\% lower limit on the scale $\Lambda$  at the LHC is shown as a function of parton center of mass energy $m_{WV}$.  The Wilson coefficient is set to be $c_{q_L}^{(3)} = 1$, and the limit is set using channels $pp\rightarrow W V\rightarrow \ell\nu q \bar{q}$ (upper two plots) and $pp\rightarrow W h\rightarrow \ell\nu b\bar{b}$ (lower plot) for integrated luminosities $L = 300 \ \text{fb}^{-1}$ (solid blue) and   $L = 3 \ \text{ab}^{-1}$ (solid black).  The dashed red  line, for $m_{WV} = \Lambda_{95\%}$,  is the condition for the consistency of weakly coupled effective field theory.  The dashed orange line, for $m_{WV} =\frac{4\pi}{g} \Lambda_{95\%}$,  is the condition for the consistency of most strongly coupled effective field theory (operator enhanced by $(4\pi)^2/g^2$). If  the limit $\Lambda_{95\%} > m_{WV} $ in a  particular $m_{WV} $ bin,  it is consistent with SM effective field theory. For $WV$ process, we have explored two benchmark values for the boosted $V$-jet tagging efficiency and the reducible background, i.e. $\epsilon_V = 0.3, n_{\rm red} = n_{\rm SM}$ (upper left) and  $\epsilon_V = 0.1, n_{\rm red} = 0$ (upper right).  In addition, we assume that the $W(V)$ polarization tagging efficiencies are $\epsilon_L = 0.75, \epsilon_T = 0.5$. }
\label{fig:reachLambda1}
\end{figure}

\subsection{Reach of the scale of new physics}

Based on our analysis of the semi-leptonic channels of di-boson production, we now turn to the reach of new physics, parameterized by the dimension-6 EFT operators in Eq.~(\ref{eq:d6ops}), through precision measurement in this channels. We make projections for the $95\%$ confidence level reach of the scale $\Lambda$, denoted as $\Lambda_{95\%}$, while setting the corresponding Wilson coefficient $c_i =1$. 

As shown in Table~\ref{tab:obs}, production of di-boson final states in the high energy limit only depends on certain combination of the EFT operators. Hence, in generating signal events, it is sufficient to include one of the operators in the combination. In particular, we generate the events using $\mO_{HW}$ operator for the combination  $c_{q_L}^{(3)}$, while for the combination $c_B + c_{HB} - c_{2B}$, we use operator $\mO_{HB}$. We are not going to discuss  the $U(1)_Y$ current-current fermionic operators ($\mO_{R}^u, \mO_{R}^d, \mO_{L}^q$), as the sensitivity to  them is expected to be similar to that of $\mO_{HB}$. 
 We ``turn on" one operator at a time. Including multiple operator at the same time can lead to potential correlations and flat directions. We will leave a more comprehensive treatment for a future study. We first show the bound from $WV, Wh$ channels in each di-boson invariant mass (or equivalently parton center of mass energy) bin in Fig.~\ref{fig:reachLambda1} 
for integrated luminosities $L = 300 \text{fb}^{-1}$ or $L = 3 \text{ab}^{-1}$.  For the  studies of semi-leptonically decaying channel of  $WV$ by  CMS at 8 TeV 
\cite{Sirunyan:2017bey},  the systematics is dominated by the $W+jets$ background normalization, which is around 20\%. We expect that significant improvement in the HL-LHC, and the systematics can be reduced. Similar expectations apply  to the $Vh$ channels. In making this figure, we have assumed that the systematical error is $5\%$. For our final combined results presented later, we vary the systematics between $3\%$ and $10 \%$. 

For the $WV$  channel, in each di-boson invariant mass bin, we divided the  partonic scattering angle $\cos\theta_V$ into four bins $[0,0.2],[0.2,0.4],[0.4,0.6],[0.6,1.0]$. Then, we combined the bins with number of event greater than 5. This effectively put a cut on $\cos\theta_V$ which enhances the longitudinal new physics signal. From Fig.~\ref{fig:reachLambda1}, we can see that higher energy bins, or equivalently larger $m_{WV}$ or  $m_{Wh}$ bins,  generically yield better reaches. This is due to the inclusion of the systematical error, which limits the effectiveness of lower energy bins.   For the high-luminosity LHC ($L = 3 \ \text{ab}^{-1}$),  the reach of the cut-off $\Lambda_{95\%}$ in each di-boson invariant mass bin is larger than the value of $m_{WV}(m_{Wh})$. Therefore, the reach is  consistent with effective field assumptions from integrating out weakly coupled UV physics with $c_{q_L}^{(3)} \sim 1$. On the other hand,  for integrated luminosity $L = 300\ \text{fb}^{-1}$, not all the bins can be used to put consistent bound for the $\Lambda$ in the weakly coupled theory~\cite{Biekoetter:2014jwa,Contino:2016jqw}. It is still useful when the new physics is strongly coupled and the Wilson coefficients are enhanced by the strong coupling, as will be discussed in the Section~\ref{sec:reachnp}.  In Fig~\ref{fig:reachLambda1},  we have also plotted the limit on the validity of EFT in most strongly coupled case $c_{q_L}^{(3)} \sim \frac{(4\pi)^2}{g^2}$ (orange dashed line), which can arise if $q_L$ is fully composite.

\begin{figure}[!tb]
\begin{center}
\includegraphics[width=0.49\textwidth]{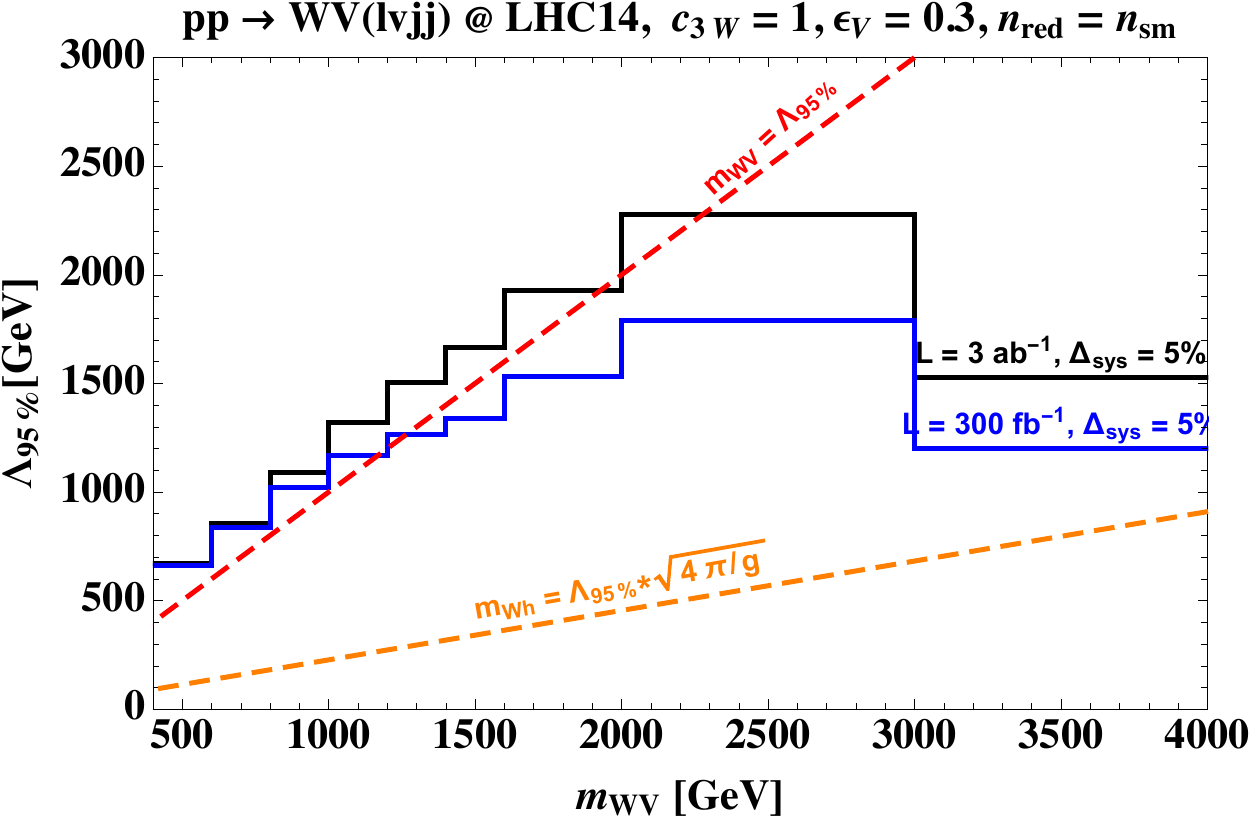}
\includegraphics[width=0.49\textwidth]{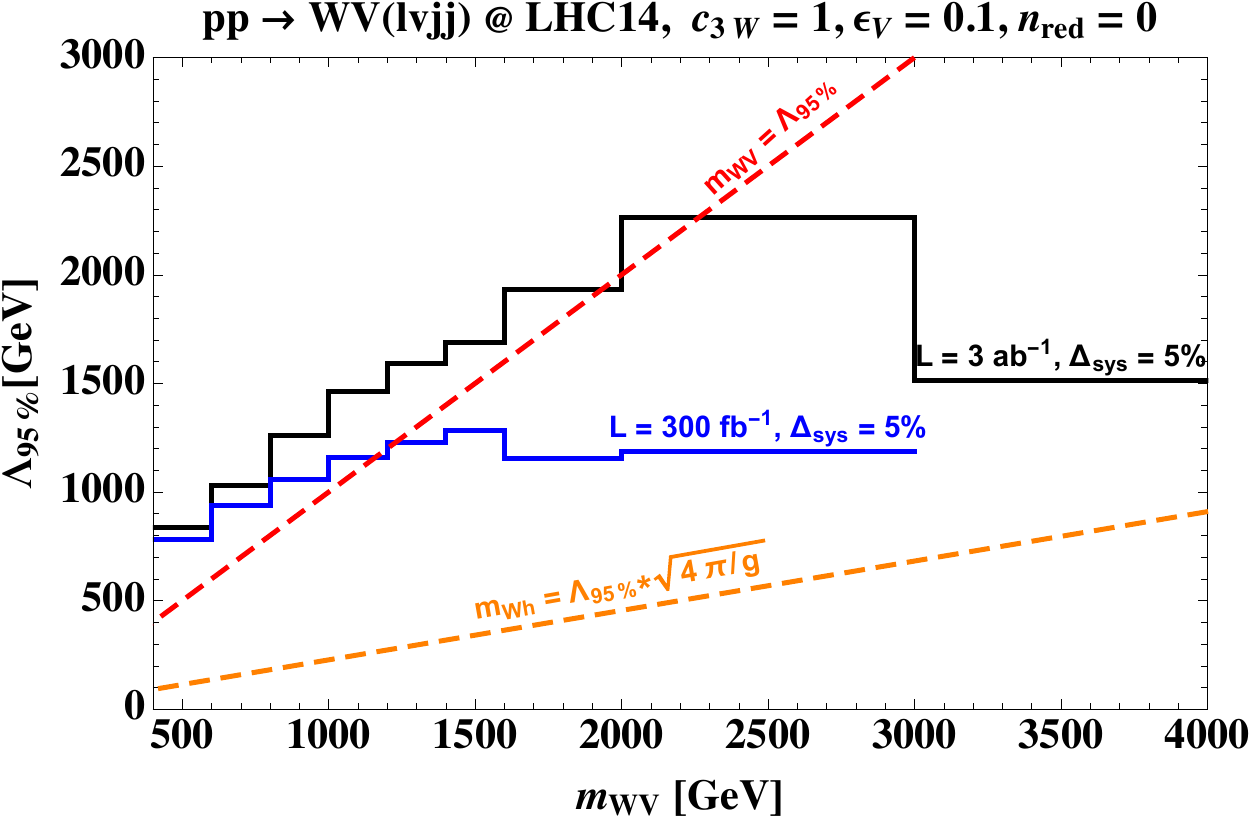}
\\
\end{center}
\caption{95\% lower limit on the scale $\Lambda$ at the LHC   for Wilson coefficient $c_{3W} = 1$ in the channel $pp\rightarrow W V\rightarrow \ell\nu q \bar{q}$ , with systematics $\Delta_{sys} = 5\%$, for the integrated luminosities $L = 300 \ \text{fb}^{-1}$ (solid blue) and   $L = 3 \ \text{ab}^{-1}$ (solid black). The dashed red  line, for $m_{WV} = \Lambda_{95\%}$,  is the condition for the consistency of weakly coupled effective field theory (Wilson coefficient is $\mO(1)$).  The dashed orange line, for $m_{WV} = \sqrt{\frac{4\pi}{g}} \Lambda_{95\%}$,  is the condition for the consistency of most strongly coupled effective field theory (Wilson coefficient is $4\pi/g$). }
\label{fig:reachLambda2}
\end{figure}

We have also evaluated the reach on $\mO_{3W}$  using semi-leptonically decaying $WV$ channel. The result is shown in Fig.~\ref{fig:reachLambda2}, where we have performed an analysis similar to the $\mO_{HW}$ case, i.e., using similar $\cos\theta_V$ bins and assumption about the reducible backgrounds. 
 As expected, the sensitivity to the $\mO_{3W}$ operator  is weaker than the  $\mO_{HW}$ operator.  This is due to the fact that the new signal from the the $\mO_{3W}$ operator does not interfere with SM amplitudes (see Table~\ref{tab:obs} and also Ref.~\cite{Azatov:2016sqh}). In fact, it only contributes to di-boson states with helicities $\pm \pm$\footnote{In principle, we can use the correlation between  $\cos\theta^*$ distribution of the two gauge bosons to distinguish the $++$ and $+-$ helicity final states. But for the semi-leptonically decaying channel, since we can't distinguish the up-type quark with down-type quark for the $W$-boson decay, so the information will be lost.}. The corresponding SM amplitudes with same helicities  go to zero in the high energy limit, scaling like $m_W^2/E^2$. From Fig.~\ref{fig:reachLambda2}, we can infer that the reach is  in mild tension with weakly coupled effective field theory even for the high-luminosity LHC. But for strongly coupled transverse gauge bosons, the Wilson coefficient can be enhanced by the strong coupling. In this case, the projected reach is consistent with effective field theory as long as the coupling is large enough (see the orange dashed line in Fig.~\ref{fig:reachLambda2} for the most strongly coupled case with $c_{3W} \sim 4\pi/g$).
This reach maybe further improved by using azimuthal angle distribution of the decay information of the $W,Z$ bosons, which results in interference with leading non-vanishing SM amplitude (see Ref.~\cite{Azatov:2017kzw,Panico:2017frx}). We will  not explore this possibility further here.

\begin{figure}[!h]
\begin{center}
\includegraphics[width=0.95\textwidth]{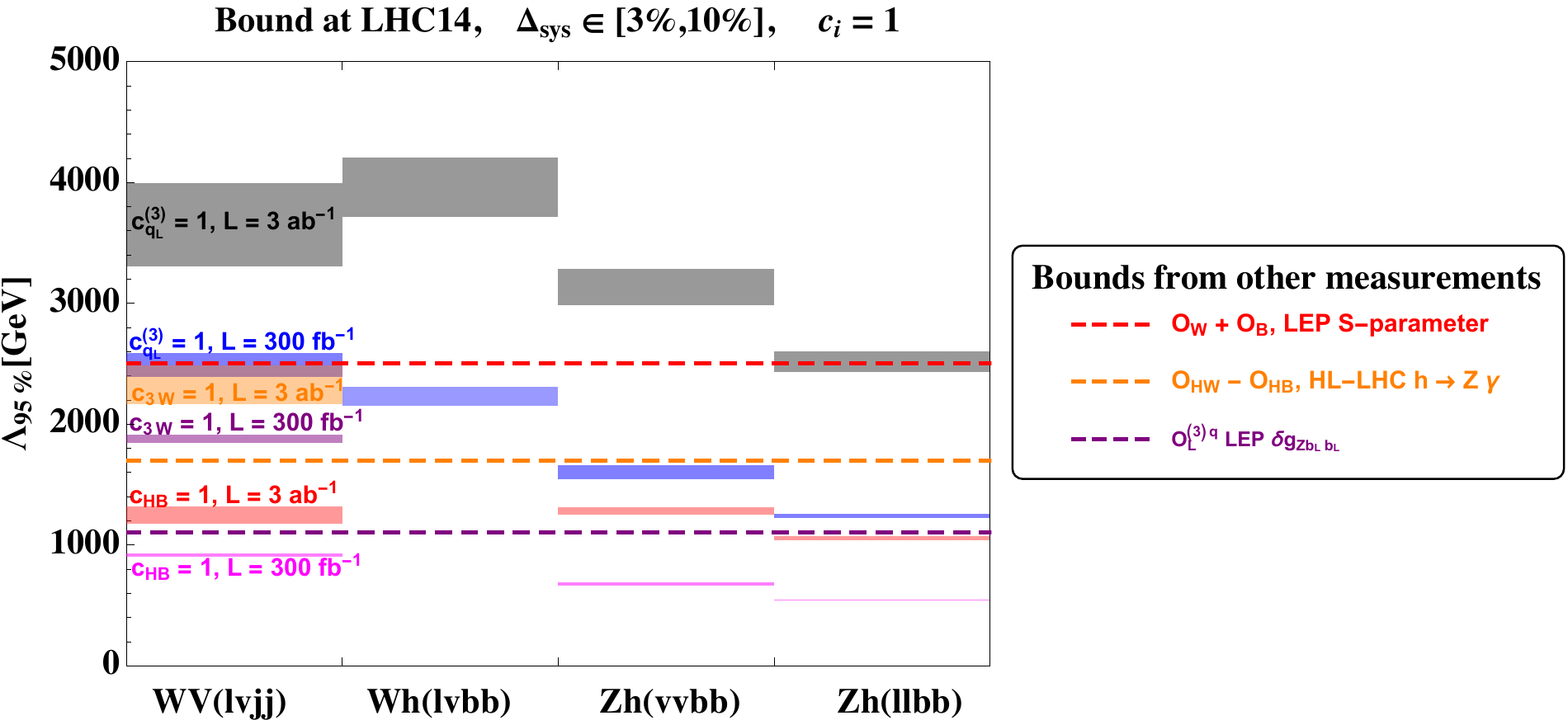}
\end{center}
\caption{Reach in different channels at the  14 TeV LHC for different combinations of the operators assuming the systematical error varying from 3\% to 10\%. The grey and  blue regions  denote the reach of the scale in the case of  $c_{q_L}^{(3)} = 1$  for integrated luminosities $L = 3 \ \text{ab}^{-1}$ and  $L = 300 \ \text{fb}^{-1}$, respectively.  The red and  magenta regions denote the reach in the case of $c_{B} + c_{HB} - c_{2B} = 1$  for integrated luminosities $L = 3 \ \text{ab}^{-1}$ and  $L = 300 \ \text{fb}^{-1}$. The orange and  purple regions denote the reach of the size of $\mO_{3W}$ operator with $c_{3W}= 1$,  for integrated luminosities $L = 3 \ \text{ab}^{-1}$ and  $L = 300\  \text{fb}^{-1}$. We also show the  present bound from LEP S-parameter on the combination of operators $\mO_{W}$ and $\mO_{B}$ with $c_{W} + c_{B} = 1$ (red dashed line),  the bound from LEP  $\delta g_{Zb_L \bar{b}_L}$ measurement on the operator $c_{L}^{(3)q} = 1/4$ (purple dashed line), based on flavour-universal effects. The case of  $c_{HW}- c_{HB} = 1$  is shown in orange dashed line from  $3 \ \text{ab}^{-1}$  HL-LHC measurement of $h\rightarrow Z\gamma$ decay partial width, with a projected precision of $\sim 20\%$ from Ref.~\cite{Dawson:2013bba}.}
\label{fig:LHCreach}
\end{figure} 

Finally, we combine all the bins and make projections on the reach of cut-off $\Lambda$ for different operators in different processes. The results are summarized in Fig.~\ref{fig:LHCreach}. We have varied the systematics from $3\%$ to $10\%$. For the semi-leptonically decaying $WV$ channel, we only show the benchmark values for $\epsilon_V^{\rm tag} = 0.3, n_{\rm red} = n_{\rm irred}$. For the second benchmark point of \Eq{eq:benchmark}, there is no big difference except the dependence on the systematic uncertainty is weaker. This is because of the assumption of zero reducible background. From Fig.~\ref{fig:LHCreach}, we can infer that for the case of $c_{q_L}^{(3)} = 1$, the most important bound comes from both $Wh(\ell \nu b\bar{b})$ and $WV (\ell\nu jj)$ channels.  Taking $\Delta_{sys} = 5\%$ as a benchmark point, the reaches in these two channels  are comparable, around 3.8(2.5) TeV in the $WV(\ell\nu jj)$ channel for integrated luminosities $L = 3 \ \text{ab}^{-1}(300 \ \text{fb}^{-1})$, and 4.0(2.3) TeV for the $Wh$ channel. 
Note that $c_{q_L}^{(3)}$ is the combination of the operators $c_{q_L}^{(3)} = c_W + c_{HW}  - c_{2W}+ 4 c_L^{(3)q}$. If we assume that there is no big cancellation in different Wilson coefficients, we can compare the reach from Di-boson processes with the bound from EWPT at the LEP and Higgs coupling measurement at the HL-LHC, even though the later two depend on different combinations of operators (see Table~\ref{tab:obs} ). The operator $\mO_W$ will contribute to the $S$-parameter~\cite{Peskin:1991sw,Barbieri:2004qk}. Suppose it is the dominant contribution,  the bound is  $\sim$ 2.5 TeV at 95\% CL for $c_W = 1$. $\mO_{HW}$ will contribute to the Higgs rare process $h\rightarrow Z\gamma$. The  $h\rightarrow Z\gamma$ measurement at HL-LHC will put a limit around 1.7 TeV~\cite{Dawson:2013bba} for $c_{HW} = 1$. For the flavour-universal operator $\mO_L^{(3)q}$, from LEP $\delta g_{Zb_Lb_L}$ measurement,  the bound is around $1.1$ TeV  for $c_L^{(3)q} = 1/4$ \cite{Pomarol:2013zra,Ellis:2014jta}\footnote{ $c_L^{(3)q} = 1/4$ is chosen such that $c_{q_L}^{(3)} =1$ (see \Eq{eq:comb}).}.  We have shown the three bounds as the red, orange, purple dashed lines  in Fig.~\ref{fig:LHCreach}.  The comparison above shows diboson measurement is very promising to probe the new physics scenario in which the operators considered here give the most important effect.   For the operator $\mO_{2W}$, it will contribute to the four fermion operator by equation of motion, especially it will contribute to Drell-Yan processes $q\bar{q} \rightarrow \ell^+\ell^-$. This has been studied in Ref.~\cite{Farina:2016rws} and the expected reach is 13.4 TeV at the HL-LHC for $c_{2W} = 1$. Usually, this operator will be suppressed by a factor of $g^2/g_*^2$. However,  in certain scenario with  strong multi-pole interactions (the so called Remedios scenario) in Ref.~\cite{Liu:2016idz}, this operator may become as relevant as others. We will discuss this in detail in the next section. For the operator combinations of $c_B + c_{HB} -c_{2B}$, the reach is relatively weak (1.3 TeV at the HL-LHC ) from di-boson process. This is  a result of the smallness of the hyper-charge coupling $g^\prime$. This makes it difficult to compete with S-parameter and $hZ\gamma$ measurement, and the reach is also not consistent with weakly coupled effective field theory. We finally mention that the bound for $\mO_{3W}$ is 2.4 (1.9)TeV at the  3 ab$^{-1}$ (300 fb$^{-1}$), which is also only meaningful if its Wilson coefficient is enhanced by a strong coupling.

\section{Reach of new physics scales in different scenarios}
\label{sec:reachnp}
In Section~\ref{sec:semileptonic}, we have presented the projection on the reach of $\Lambda$ in an model independent way with unit Wilson coefficients ($c_{q_L}^{(3)} = 1, c_{3W} = 1$ etc).  In different new physics scenarios, the size of Wilson coefficients can be quite different. Assuming that the new physics is broadly characterized by a mass scale of new states $m_*$ and a coupling $g_*$, the Wilson coefficients are 
\beq
c_i \sim \frac{g_*^n}{g_{\rm SM}^n}\left( \frac{g_*^2}{16\pi^2}\right)^{n_{\rm Loop}}, \qquad n \leq n (\mO_i) - 2
\eeq
where $g_{\rm SM}$ denotes the SM gauge and  Yukawa couplings $g,g^\prime, y_f$. $n(\mO_i)$ is the number of fields in the operator $\mO_i$. In particular, $n(\mO_{W,B,HW,HB, 3W}) = 3, n(\mO_{2W,2B} ) = 2, \text{and } n (\mO_{L,R}^q) = 4$. Note that a covariant derivative is not counted as  a field. $n_{\rm Loop}$ is the number of loops needed to generate the operator. Note that $n$ can also be negative. The bounds $\Lambda_{95\%}$ obtained in the previous section can be easily translated into the bounds on the mass scale $m_*$,  as functions of  Wilson coefficients $c_i$:
\beq
m_* > \Lambda_{95\%}^{c_ i = 1} \times \sqrt{c_i}
\eeq
 In the following, we will 
consider the Strong-Interacting-Light-Higgs (SILH), strong multi-pole interaction (Remedios), and the (partially) composite fermion scenarios.

\subsection{SILH scenario}
We  start with the SILH scenario~\cite{Giudice:2007fh}. There are two basic assumptions.  First, the  Higgs  and the longitudinal components of the SM gauge bosons are pseudo-Nambu-Goldstone-bosons associated with the global symmetry breaking in a strongly interacting sector~\cite{Agashe:2004rs}. In addition,  the SM fermions acquire masses from their linear mixing with corresponding strongly interacting sector states (the so-called partial compositeness~\cite{Kaplan:1991dc}). This leads to the following power-counting rules for the Wilson coefficients:

\begin{itemize}
\item  Each Higgs and Goldstone fields will be associated with a strong coupling $g_*$ in the operators which preserve the global symmetries of the strongly interacting  sector, including those which are non-linearly realized.
\item Explicitly breaking of the strongly interacting sector symmetries will be associated with SM gauge couplings and Yukawa couplings, $g$, $g^\prime$, and   $y_f$.
\end{itemize}
Following these rules, we have summarized the size of the Wilson coefficients of the operators for the SILH scenario in the second row of Table~\ref{tab:wilsoncoeff}. None of the operators considered in our paper is enhanced by  the strong coupling $g_*$, mainly due to the fact that the transversely polarized gauge bosons belong to the elementary sector.  In the second row  of Table~\ref{tab:bounds}, we summarize the reach of the mass scales in the SILH scenario  from HL-LHC measurements of  di-boson, $h\rightarrow Z\gamma, h\rightarrow \gamma\gamma$~\cite{Dawson:2013bba}, and di-lepton processes.  For comparison, we have also included the bound from  $S$-parameter measurement.  In comparison with other measurements,  di-boson processes have the best reach in the SILH scenario. 

 \begin{table}[!h]
\renewcommand{\arraystretch}{1.2}
{
\begin{center}
\begin{tabular}{|c|c|c|c|c|c|c|c|c|c|}
\hline 
Model &${\cal O}_{2W}$ & ${\cal O}_{2B}$& ${\cal O}_{3W}$ & ${\cal O}_{HW}$ & ${\cal O}_{HB}$ & ${\cal O}_{W,B}$ & ${\cal O}_{BB}$  \\
\hline
{\footnotesize SILH }& $\frac{g^2}{g_*^2}$  & $\frac{g^{\prime 2}}{g_*^2}$ &$\frac{g^2}{16\pi^2}$& $\frac{g^2_*}{16\pi^2}$   & $\frac{g^2_*}{16\pi^2}$&   $1$ &$\frac{g^2}{16\pi^2}$  \\
{\footnotesize Remedios } & 1 & 1& $\frac{g_*}{g}$&  &  &  &     \\
{\footnotesize  Remedios+MCHM  }& 1&  1&$\frac{g_*}{g}$& $1$       & $1$ &
$1$ & 1 \\
{\footnotesize  Remedios+$ISO(4)$  } &1 & 1& $\frac{g_*}{g}$ & $\frac{g_*}{g}$ & $1$   &  $1$   & 1   \\
\hline
\end{tabular}
\end{center}
}
\vspace{0.3cm}
\caption{Power counting of the size of the Wilson coefficients in different scenario, where $g_*$ denotes the coupling in the strong sector. For completeness, we have added $\mO_{BB} = g^{\prime 2} H^\dagger H B_{\mu\nu} B^{\mu\nu}$. }
\label{tab:wilsoncoeff}
\end{table}%

 \begin{table}[!h]
\renewcommand{\arraystretch}{1.2}
{
\begin{center}
\begin{tabular}{|c|c|c|c|c|c|c|c|c|c|}
\hline 
Model  & Di-boson & S-parameter & LHC $h\rightarrow Z \gamma$ & LHC $h\rightarrow \gamma\gamma$ & LHC dilepton  \\
\hline
{\footnotesize SILH }& 4.0 & 2.5 & $ 1.7 \sqrt{\frac{g_*}{4\pi}}$ & 0.34 & $0.69\sqrt{\frac{4\pi}{g_*}}$  \\
{\footnotesize Remedios } & $ 10.6 \sqrt{\frac{g_*}{4\pi}}$ &  & & & 13.4    \\
{\footnotesize  Remedios+MCHM  }&$ 10.6 \sqrt{\frac{g_*}{4\pi}}$  &  2.5& 1.7& 6.5& 13.4    \\
{\footnotesize  Remedios+$ISO(4)$  } & $ 17.6 \sqrt{\frac{g_*}{4\pi}}$&2.5 &$ 7.5 \sqrt{\frac{g_*}{4\pi}}$ & 6.5&13.4    \\
\hline
\end{tabular}
\end{center}
}
\caption{ The bounds (in TeV) for different scenarios from different measurements. The LHC measurement are prospectives at the integrated luminosity $L = 3\,\text{ab}^{-1}$.  }
\label{tab:bounds}
\end{table}%

\subsection{Strong multi-pole interaction (Remedios) scenario}
Ref.~\cite{Liu:2016idz} considers the possibility that the SM transverse gauge bosons are part of the strong dynamics. This so called Remedios scenario is based on the observation that the normal SM gauge interactions (mono-pole) and multi-pole interactions (involving the field strength and its derivatives) have  different symmetry structure. Therefore, they can have different coupling strengths in principle. The small Standard Model couplings, such as $g$,  control the renormalizable interactions between the gauge boson and the fermions. At the same time,  the large coupling $g_*$ determines the strength of the multi-pole interactions of the gauge bosons with the resonances of the strong sector. This will lead to the following new power-counting rules for the gauge bosons:
\begin{itemize}
\item The field strengths of the gauge boson and their derivatives are  associated with a strong coupling $g_*$,  if the interactions preserve the global symmetries of the strong sector. The normal gauge interactions are realized by  changing the partial derivative to covariant derivative: $\partial_\mu \rightarrow D_\mu = \partial_\mu -i  gA_\mu$.
\end{itemize}
In this case, the $\mO_{3W}$  operator is enhanced by the strong coupling, while the $\mO_{2W,2B}$ operators have $\mO(1)$ Wilson coefficients. The power counting of these operators considered in this scenario have been summarized in the third row  of Table~\ref{tab:wilsoncoeff}.
We can consider further the scenarios that both transverse gauge bosons and Higgs bosons are part of the strong dynamics. Depending on the symmetry of the strong sector, we have two benchmark scenarios:
\begin{itemize}
\item \text{Remedios + MCHM}: the symmetry breaking of the strong sector will be $SO(5)\times \widetilde{SU(2)} \times U(1)_X \rightarrow SO(4)\times \widetilde{SU(2)} \times U(1)_X$, where another global symmetry $\widetilde{SU(2)}$ is needed to stablize the Higgs potential. 
\item \text{Remedios} +$ ISO(4)$: the symmetry breaking of the strong sector will be $ISO(4)\times  U(1)_X \rightarrow SO(4) \times U(1)_X$, where the $ISO(4)$ is the non-compact group $SO(4)\rtimes T^4$.
\end{itemize}
The corresponding power-counting rules for the size of the Wilson coefficients are presented in the fourth and fifth rows  of Table~\ref{tab:wilsoncoeff}.  We summarize the reaches for these three benchmark scenarios from different measurements of Table~\ref{tab:bounds}. Several comments are in order. If only the field strengths are strongly coupled (3rd row), the most relevant operators are $\mO_{2W}$  with $\mO(1)$ Wilson coefficients and $\mO_{3W}$ with enhanced Wilson coefficient $\sim \mO(g_*/g)$. Di-lepton measurements at HL-LHC will reach 13.4 TeV. The reach from  Di-boson measurements are  weaker, which is 10.6 TeV for the most strongly interacting case $g_* = 4 \pi$.  The projection is similar for the $Remedios$ + MCHM scenario.  For the$Remedios + ISO(4)$ scenario, $\mO_{HW}$ is enhanced by the strong coupling $g_*$. Its Wilson coefficient is $g_*/g$. As a result, Di-boson measurement can reach higher $\sim 17.6 \sqrt{g_*/4\pi}$ TeV,  which becomes better than Di-lepton measurement for large coupling $g_* \gtrsim 7$.

\subsection{Partially Composite fermions}

Finally, we discuss the fermionic  operators. We focus on the operators $\mO_L^{(3)q}$, and we expect the conclusions for other fermionic operators are similar. We will also focus on the flavor-universal  effects, that are invariant under $SU(3)$ flavor transformation. Other effects will be suppressed by the Yukawa couplings under the assumption of minimal flavor violation (MFV) \cite{DAmbrosio:2002vsn}.
As discussed before, the LHC Diboson measurement ($\Lambda^{c_L^{(3)q} = \frac14}_{95\%}\sim$ 4 TeV) will be much better than the LEP measurement $\Lambda^{c_L^{(3)q} = \frac14}_{95\%}\sim$ 1.1 TeV) for such effects. Now if we assume that  the SM fermions have some degrees of compositeness $\epsilon_{q_L}$ (for example partial compositeness scenario in \Ref{Kaplan:1991dc}),  the  size of Wilson coefficient of the $\mO_L^{(3)q}$ by power counting  is:
\beq
c_{L}^{(3)q_L} \sim  \frac14 \frac{g_*^2}{g^2} \epsilon_{q_L}^2
\eeq
where we have factored out a $1/4$ factor to be consistent with above consideration. The HL-LHC Di-boson measurement will reach the mass scale:
\beq
m_* \gtrsim 77 \frac{g_*}{ 4 \pi} \epsilon_{q_L}\, \TeV \qquad  @ 95\% \text{CL}
\eeq
In the meantime, the following four-fermion operator will also be present in the low energy effective field theory:
\beq
\label{eq:L4f}
\mL_{4f} =  \frac{g_*^2 \epsilon_{q_L}^4}{m_*^2 }  \bar{q}_L \gamma^\mu q_L\bar{q}_L \gamma_\mu q_L
\eeq
This will lead to energy growing behaviour in the di-jet processes at the LHC. 
The present  bound from ATLAS di-jet measurement~\cite{ATLAS:2016lvi} at the 13 TeV with the integrated luminosity of $15.7\text{fb}^{-1}$ is given by (see Ref.~\cite{Bellazzini:2017bkb,Domenech:2012ai}):
\beq
m_* \gtrsim 62 \frac{g_*}{ 4 \pi} (\epsilon_{q_L})^2\, \TeV \qquad  @ 95\% \text{CL}
\eeq
\Ref{Alioli:2017jdo} has studied the prospectives  on the following operator at the HL-LHC with 13 TeV center-of mass energy\footnote{Actually, this operator arises from  $-\frac{1}{2}(D_\mu G^{A\mu\nu})^2$ by equation of motion of the gluon fields.}:
\beq
  -\frac{g_*^2 \epsilon_{q}^4}{ 2m_*^2 } \left( \sum_q\bar{q} \gamma^\mu  T^A   q  \right)^2
\eeq
where $T^A$ is the generators of QCD $SU(3)_c$ group. The expected $95\%$ CL bound on the scale is
\beq
m_* \gtrsim 83 \frac{g_*}{ 4 \pi} (\epsilon_{q_L})^2\, \TeV \qquad  @ 95\% \text{CL}
\eeq
Although this operator is different from the one in \Eq{eq:L4f}, it can provide a rough idea about what the scale is probed  in the di-jet process at the HL-LHC.
 We can see that for the smaller values of $\epsilon_{q} < 0.9$, the LHC Di-boson measurement can be more promising than the di-jet process. 

\section{Conclusions}
\label{sec:conclusion}

The future runs of the LHC in the next decade or so will collect nearly thirty times more data than currently available. There is great potential to improve the precision measurements with this new data set. The measurements with SM electroweak sector is particularly important, as it is closely related to new physics associated with electroweak symmetry breaking. Studies of Di-boson channels, $VV$ and $Vh$ where $V$ can be SM $W$ and $Z$, give a promising window into such new physics. Such measurements can be complementary to the direct search of new physics particles. In certain scenarios, new physics particles can be too heavy to be produced at the LHC. At the same time, their presence can lead to observable effects in precision measurements. 

In this paper, we parameterize the new physics effects with dimension 6 EFT operators. We focus on operators which are most relevant for the di-boson final states. In particular, we study the reach in the semi-leptonic final states. In order to fully take advantage of the larger effect of EFT operators at higher energies, we need to select final states which interfere with the SM background. While this is guaranteed for the $Vh$ channel, we have to select longitudinally polarized $W$ and $Z$ in the $WW$ and $WZ$ channels. There are two possible strategies to achieve polarization tagging. First, the angular distributions of longitudinally and transversely polarized gauge bosons are different. This effect is most dramatic in the $WZ$ final state with the so called amplitude zero in the central region for the transverse vector bosons. This has been crucial for the analysis in the pure leptonic channel~\cite{Franceschini:2017xkh}. For the semi-leptonic channel we studied here, since we can not distinguish hadronic $W$ and $Z$ very well, this effect is less prominent. Another approach is to directly tag the polarization of the gauge boson by the angular distribution of their decay products. In our study, we use a combination of both approaches. Since the precision measurements typically focus on cases where $S/B$ is small, the sensitivity depends crucially on systematic error and background estimates (in particular reducible background). For the reducible background of semi-leptonic $WV$ channel, we have considered the dominant background $W$+jets at parton level and applied the $W$-tagging efficiency and QCD-jet mis-tagging efficiency based on the study of  Ref.~\cite{ATLAS:2017jiz}. The resulting two benchmarks are summarized in \Eq{eq:benchmark}. For the $Vh(b\bar{b})$ channel, we have adopted the study of \Ref{Tian:2017oza} about the reducible backgrounds in the 0,1, and 2 lepton channels, which leads to \Eq{eq:vhbenchmark} as our benchmark in these channels. Our results shows that precision measurement at the LHC can have good sensitivity in probing new physics at multiple-TeV scale. It can surpass the sensitivity of LEP precision measurements, such as those from the S-parameter and $Z$ coupling measurements. Compared with fully leptonic decaying $WZ$ channels, the semi-leptonic decay $WV$ channel has order of magnitude larger rate. At the same time, semi-leptonic decay channels suffer from large reducible backgrounds. During the up coming runs of the LHC, we expect significant improvement in understanding the reducible background and reducing systematics.
Anticipating this, we make optimistic projections of the reducible background for  the semi-leptonic decay channels based on extrapolations of ATLAS study.   Based on this, our result ($\Lambda^{c_{q_L}^{(3)}}_{95\%} \sim $ 4 TeV)  is better than the fully leptonic $WZ$ channel ($\Lambda^{c_{q_L}^{(3)}}_{95\%} \sim $ 3.2 TeV) studied in \Ref{Franceschini:2017xkh}. As an application of our result, we derived the reach of  the new physics scale in several new physics scenarios. In the SILH scenario, which models the generic feature of composite Higgs models, the diboson measurement can be more sensitive than other experimental observables. For the scenario with strong multiple interactions (the so called Remedios), di-boson is either slightly weaker or comparable with the measurements in di-lepton channel. 

It is worth emphasizing that the estimates we made here are based on our assumptions about systematics and efficiencies achievable at the HL-LHC. More detailed and realistic studies, presumably based on real data and full fledged simulations, would be necessary to determine the precise reach. In this sense, the numbers presented here are better considered as benchmarks or targets, which could give us good reach in these channels. We have also identified several directions in which improvements can be crucial to enhance the sensitivity in the di-boson channel. Obviously, any new technique to tag the polarization of the vector bosons can be very helpful. A major direction to pursue is the tagging of  polarization of the hadronic $W$ and $Z$. In addition, distinguishing hadronic $W$ and $Z$ can also be very helpful in enhancing longitudinal final states.

\section{Acknowledgement}
We are grateful to Andrea Tesi for numerous helpful discussion and collaboration during the early stages of this work. We would also like to thank Francesco Riva, Andrea Wulzer for helpful discussions and thank Nurfikri Norjoharuddeen, Oliver Majersky and Ece Akilli  for bringing \Ref{ATL-PHYS-PUB-2015-033} to our attention. LTW is supported by the DOE grant DE-SC0013642. DL is supported in part by the U.S. Department of Energy under Contract No. DE-AC02-06CH11357.

\appendix

\section{Cross sections of the Di-boson processes at the LHC}
 In Table~\ref{tab:xspt}, we have reported the cross section for the di-boson processes at the 14TeV LHC as a function of the cut-off $\Lambda$ in each $p_T$ bin with the Wilson coefficient $c_{HW}$ setting to one. The cross sections are calculated using MadGraph \cite{Alwall:2014hca} at LO simulation with parton level cuts $|\eta_{W,Z,h}| < 2.5$. We can see clearly that the new physics effects manifest in the purely longitudinal helicity final states of $W,Z$ gauge bosons and the Higgs boson with energy growing behaviour. It results in the fact that the coefficients of $1/\Lambda^2$ become larger as the $p_T$ increases. In addition, the ratios between the coefficients of $1/\Lambda^4$ and that of $1/\Lambda^2$ grow as the $p_T$ becomes larger, which indicates that the larger value of $\Lambda$ is needed for $1/\Lambda^2$ terms dominate.  
 For the processes including one transverse gauge bosons and one longitudinal gauge bosons (including the Higgs boson), the cross sections are comparable to the transverse one  in the low $p_T$ bin $[0,400]$ GeV and decrease  very fast as $p_T$ increases. It results in below $5\%$ of $LL$ one for the $WW,WZ$ processes and below $1\%$ of $V_L h$ for the $Vh$ processes for the largest $p_T$ bin.
 For the SM $WW$, the  purely transverse helicity  final states $TT$  dominate over $LL$ by a factor of 16 in the moderately boosted region $p_T \in [200,400]$ GeV and a factor of 12 in the highly boosted region  $p_T \in [1000,1500]$ GeV. While for the $WZ$ process, the $TT$ cross section  is only a factor of 3 of the $LL$ one in the moderately boosted region and becomes comparable to $LL$ in the highly boosted bin. This is due to amplitude zero in this process as discussed in the main text. 
 
\begin{savenotes}
\begin{table}[!ht]
\small
\begin{center}
\caption*{ }
\begin{tabular}{|c|c|c|c|c|c|c|c|}
\hline
 $\sigma$ [fb], $p_T$ [GeV] & [0,200]   & [200, 400] & [400,600]   \\
    \hline
    $W^\pm_L Z_L$  & $784 \left(1 +\frac{0.116}{\Lambda^2}+ \frac{0.00625}{\Lambda^4}\right)$& 58.5 $ \left(1 +\frac{0.682}{\Lambda^2}+ \frac{0.141}{\Lambda^4}\right)$ & 4.84 $ \left(1 +\frac{2.09}{\Lambda^2}+ \frac{1.24}{\Lambda^4}\right)$  \\
    \hline
    $W^\pm_{L(T) } Z_{T(L)}$  &$1614 \left(1 +\frac{0.0610}{\Lambda^2}+ \frac{0.00181}{\Lambda^4}\right)$ & 23.0 $\left(1 +\frac{0.419}{\Lambda^2}+ \frac{0.0611}{\Lambda^4}\right)$ & 0.598 $\left(1 +\frac{1.40}{\Lambda^2}+ \frac{0.623}{\Lambda^4}\right)$  \\
     \hline 
    $W^\pm_{T} Z_{T}$  & 5755 & 164& 12.0  \\
        \hline
        \hline
    $W^+_L W^-_L$  & $1416 \left(1 +\frac{0.0318}{\Lambda^2}+ \frac{0.00203}{\Lambda^4}\right)$& 34.0 $ \left(1 +\frac{0.597}{\Lambda^2}+ \frac{0.121}{\Lambda^4}\right)$ & 2.75 $ \left(1 +\frac{1.83}{\Lambda^2}+ \frac{1.05}{\Lambda^4}\right)$ \\
         \hline 
 $W^+_{L(T)} W^-_{T(L)}$ & $4866 \left(1 +\frac{0.00758}{\Lambda^2}+ \frac{0.000489}{\Lambda^4}\right)$& 34.6 $ \left(1 +\frac{0.130}{\Lambda^2}+ \frac{0.0207}{\Lambda^4}\right)$ & 0.848$ \left(1 +\frac{0.429}{\Lambda^2}+ \frac{0.213}{\Lambda^4}\right)$ \\
              \hline
    $W_{T}^+ W^-_{T}$  & 17987 & 523 & 39.1\\
    \hline
            \hline
    $W^\pm_L h$  & $387 \left(1 +\frac{0.149}{\Lambda^2}+ \frac{0.00776}{\Lambda^4}\right)$& 46.5 $ \left(1 +\frac{0.712}{\Lambda^2}+ \frac{0.148}{\Lambda^4}\right)$ & 4.30 $ \left(1 +\frac{2.13}{\Lambda^2}+ \frac{1.24}{\Lambda^4}\right)$ \\
         \hline 
 $W^\pm_{T} h$ & $270 \left(1 +\frac{0.0302}{\Lambda^2}+ \frac{0.000146}{\Lambda^4}\right)$& 4.93 $ \left(1 +\frac{0.0287}{\Lambda^2}+ \frac{0.000217}{\Lambda^4}\right)$ & 0.140$ \left(1 +\frac{0.0271}{\Lambda^2}+ \frac{0.000275}{\Lambda^4}\right)$ \\
             \hline
             \hline
    $Z_L h$  & $198 \left(1 +\frac{0.134}{\Lambda^2}+ \frac{0.00731}{\Lambda^4}\right)$& 24.5 $ \left(1 +\frac{0.628}{\Lambda^2}+ \frac{0.136}{\Lambda^4}\right)$ & 2.24 $ \left(1 +\frac{1.90}{\Lambda^2}+ \frac{1.14}{\Lambda^4}\right)$\\
         \hline 
 $Z_{T} h$ & $154 \left(1 +\frac{0.0361}{\Lambda^2}+ \frac{0.000353}{\Lambda^4}\right)$& 3.30 $ \left(1 +\frac{0.0688}{\Lambda^2}+ \frac{0.00501}{\Lambda^4}\right)$ & 0.0941$ \left(1 +\frac{0.165}{\Lambda^2}+ \frac{0.0413}{\Lambda^4}\right)$  \\
    \hline
    \hline
 $\sigma$ [fb], $p_T$ [GeV]    & [600,800]    &[800,1000] & [1000,1500] \\
    \hline
    $W^\pm_L Z_L$      & 0.799 $ \left(1 +\frac{4.30}{\Lambda^2}+ \frac{4.87}{\Lambda^4}\right)$ & 0.188 $ \left(1 +\frac{6.92}{\Lambda^2}+ \frac{13.4}{\Lambda^4}\right)$ & 0.0749$ \left(1 +\frac{11.9}{\Lambda^2}+ \frac{39.1}{\Lambda^4}\right)$ \\
        \hline
       $W^\pm_{L(T)} Z_{T(L)}$  & 0.0471 $\left(1 +\frac{2.91}{\Lambda^2}+ \frac{2.60}{\Lambda^4}\right)$ & 0.00634$\left(1 +\frac{4.89}{\Lambda^2}+ \frac{7.34}{\Lambda^4}\right)$ & 0.00149$\left(1 +\frac{8.01}{\Lambda^2}+ \frac{20.6}{\Lambda^4}\right)$ \\
    \hline
       $W^\pm_{T} Z_{T}$ & 1.74& 0.357 &0.121  \\
       \hline
           \hline
$W_{L}^+ W^-_{L}$ & 0.442 $ \left(1 +\frac{3.74}{\Lambda^2}+ \frac{4.13}{\Lambda^4}\right)$     & 0.102 $ \left(1 +\frac{6.13}{\Lambda^2}+ \frac{11.4}{\Lambda^4}\right)$ & 0.0405$ \left(1 +\frac{10.4}{\Lambda^2}+ \frac{32.6}{\Lambda^4}\right)$ \\
        \hline
$W_{L(T)}^+ W^-_{T(L)}$ & 0.0652 $ \left(1 +\frac{0.888}{\Lambda^2}+ \frac{0.889}{\Lambda^4}\right)$  & 0.00873$\left(1 +\frac{1.49}{\Lambda^2}+ \frac{2.47}{\Lambda^4}\right)$ & 0.00204$\left(1 +\frac{2.43}{\Lambda^2}+ \frac{6.79}{\Lambda^4}\right)$ \\
       \hline
           $W_{T}^+ W^-_{T}$  & 5.92  & 1.28 &0.475  \\
             \hline
             \hline
$W_{L}^\pm h$ & 0.726 $ \left(1 +\frac{4.22}{\Lambda^2}+ \frac{4.83}{\Lambda^4}\right)$    & 0.169 $ \left(1 +\frac{7.15}{\Lambda^2}+ \frac{13.3}{\Lambda^4}\right)$ & 0.0671$ \left(1 +\frac{12.2}{\Lambda^2}+ \frac{38.7}{\Lambda^4}\right)$ \\
        \hline
$W_{T}^\pm h$ & 0.0112 $ \left(1 +\frac{0.0283}{\Lambda^2}+ \frac{0.000218}{\Lambda^4}\right)$  & 0.00153& 0.000364 \\
             \hline
             \hline
$Z_L h$    & 0.367 $ \left(1 +\frac{3.82}{\Lambda^2}+ \frac{4.50}{\Lambda^4}\right)$    & 0.0835 $ \left(1 +\frac{6.27}{\Lambda^2}+ \frac{12.4}{\Lambda^4}\right)$ & 0.0327$ \left(1 +\frac{11.0}{\Lambda^2}+ \frac{35.9}{\Lambda^4}\right)$ \\
        \hline
$Z_T h$  & 0.00737 $ \left(1 +\frac{0.318}{\Lambda^2}+ \frac{0.165}{\Lambda^4}\right)$ & 0.000991$\left(1 +\frac{0.523}{\Lambda^2}+ \frac{0.455}{\Lambda^4}\right)$ & 0.000231$\left(1 +\frac{0.838}{\Lambda^2}+ \frac{1.24}{\Lambda^4}\right)$  \\
\hline
\end{tabular}
\bigskip
\end{center}
\caption{  Helicity cross sections (in fb) for the di-boson  processes at the 14 TeV LHC as a function of the cut-off $\Lambda$ (in TeV) in each $p_T$ bin with the Wilson coefficient $c_{HW}$ setting to one. } 
\label{tab:xspt}
\end{table}
\end{savenotes}

\section{Polarization measurement of the W boson}
\label{app:Wpolar}
To measure the $W$ polarization, we need to study the angular distribution of its decay products. We choose the polarization axis to be the direction of the $W$ in the laboratory frame. The amplitude for the $W^+(p_W) \rightarrow l^+(p_\ell) \nu(p_\nu)$ is:
\beq
\begin{split}
\mathcal{M} =\frac{g}{\sqrt{2}} \bar{u}_L(p_\nu) \gamma^\mu \nu_L(p_\ell) \epsilon^*_\mu(p_W) 
\end{split}
\eeq
Let's start from  the rest frame of the $W^+$. We parametrize the momenta of leptons as follows:
\beq
\begin{split}
& p_\ell^* = (k, \vec{k}) = (k, k \sin\theta^* \cos\varphi^*, k \sin \theta^* \sin\varphi^*  , k \cos\theta^*),\\
&  p_\nu^* =  (k, -\vec{k}) =(k, - k \sin\theta^* \cos\varphi^*, - k \sin\theta^* \sin\varphi^*  , - k \cos\theta^*)
\end{split}
\eeq
where $k = |\vec{k}| =  m_W/2$. The general expressions for the helicity spinors are given by: 
\beq
\begin{split}
\xi_L(p_\ell^*) & = 
\left(
\begin{array}{c}
-e^{-i\varphi^*/2}\sin\frac{\theta^*}{2}\\
e^{i\varphi^*/2}\cos\frac{\theta^*}{2}\\
\end{array}
\right),\qquad
\xi_R(p_\ell^*)  = 
\left(
\begin{array}{c}
e^{-i\varphi^*/2}\cos\frac{\theta^*}{2}\\
e^{i\varphi^*/2}\sin\frac{\theta^*}{2}\\
\end{array}
\right) \\
\xi_L(p_\nu^*) & = 
\left(
\begin{array}{c}
e^{-i\varphi^*/2}\cos\frac{\theta^*}{2}\\
e^{i\varphi^*/2}\sin\frac{\theta^*}{2}\\
\end{array}
\right),\qquad
\xi_R(p_\nu^*)  = 
\left(
\begin{array}{c}
-e^{-i\varphi^*/2}\sin\frac{\theta^*}{2}\\
e^{i\varphi^*/2}\cos\frac{\theta^*}{2}\\
\end{array}
\right)
\end{split}
\eeq
and the left-handed current is as follows:
\beq
\begin{split}
\bar{u}_L(p_\nu^*) \gamma^\mu \nu_L(p_\ell^*) &= 2k (0, - e^{i\varphi^*} \cos^2\frac{\theta^*}{2} + e^{-i\varphi^*}\sin^2\frac{\theta^*}{2}, i e^{i\varphi^*} \cos^2\frac{\theta^*}{2} + i e^{-i\varphi^*}\sin^2\frac{\theta^*}{2},\sin\theta^* )^\mu
\end{split}
\eeq
By using the formulae of the  polarization vectors:
\beq
\epsilon^{+\mu} =\frac{1}{\sqrt{2}}\left(
\begin{array}{c}
0\\
1\\
i\\
0
\end{array}
\right),\qquad 
\epsilon^{-\mu} =\frac{1}{\sqrt{2}}\left(
\begin{array}{c}
0\\
1\\
-i\\
0
\end{array}
\right),\qquad 
\epsilon^{0\mu} =\left(
\begin{array}{c}
0\\
0\\
0\\
1
\end{array}
\right),
\eeq
we can easily obtain the helicity amplitudes:
\beq
\begin{split}
\mathcal{M}^+ & = -g k e^{-i\varphi^*} (1 - \cos\theta^*),\qquad \mathcal{M}^- = g k e^{i\varphi^8} (1 + \cos\theta^*), \qquad  \mathcal{M}^0 = - \sqrt{2} g k \sin\theta^*
\end{split}
\eeq
which leads to the distribution in \Eq{eq:polar}.
Turn to the laboratory frame and suppose that we can reconstruct the $z$-momentum of the neutrino by imposing the condition that the system of lepton-neutrino should correctly reproduce the mass of the $W$ boson.
 The momentum of the charged lepton and neutrino in the laboratory frame  can be obtained from the momentum in the  $W^+$ rest frame  by a Lorentz boost:
\beq
\begin{split}
 E_\ell &= \gamma \left(k + \vec{v}\cdot \vec{k}\right), \qquad \vec{p}_\ell = \vec{k} +\vec{v} \left(\gamma k   +\frac{\gamma - 1}{v^2} \vec{v}\cdot\vec{k} \right)\\
  E_\nu &= \gamma \left(k - \vec{v}\cdot \vec{k}\right), \qquad \vec{p}_\nu = -\vec{k} +\vec{v} \left(\gamma k -\frac{\gamma - 1}{v^2} \vec{v}\cdot\vec{k} \right)\\
 \end{split}
\eeq
where $\vec{v} = \frac{\vec{p}_W}{E_W}, v = |\vec{v}|$ is the velocity of the $W^+$ in the laboratory frame.
Then the formulae of $\cos\theta^*$ can be obtained by the energy difference of the lepton and neutrino in the laboratory frame as follows:
\beq
\cos\theta^* = \frac{\vec{v}\cdot \vec{k}}{k v}  = \frac{E_\ell - E_\nu}{|\vec{p}_W|}= \frac{|\vec{p}_\ell| - |\vec{p}_\nu|}{|\vec{p}_\ell +\vec{p}_\nu | }
\eeq
where we have used:
\beq
k = \frac{m_W}{2}, \qquad  |\vec{p}_W| = m_W\gamma v
\eeq

\pagestyle{plain}
\bibliographystyle{jhep}
\small
\bibliography{biblio}

\end{document}